\documentclass[11pt]{article}

\usepackage{amsmath,amssymb,cite,color,fullpage,graphicx,hyperref,tikz-cd}
\numberwithin{equation}{section}

\let\originalleft\left
\let\originalright\right
\renewcommand{\left}{\mathopen{}\mathclose\bgroup\originalleft}
\renewcommand{\right}{\aftergroup\egroup\originalright}
\newcommand{\ed}{\mathop{}\!\mathrm{d}}
\newcommand{\QE}{\text{\tiny QE}}
\DeclareMathOperator\sign{sign}

\begin{document}

\begin{titlepage}
	\begin{center}
        \hfill\vskip 0.3in
        {\LARGE\bf Photon rings in a holographic toy model}
        \vskip 0.3in

		{\large St\'ephane Detournay${}^{a}$, Sahaja Kanuri${}^{a}$, Alexandru Lupsasca${}^{b}$,\\ Philippe Spindel${}^{c,d}$, Quentin Vandermiers${}^{a}$, and Raphaela Wutte${}^{a,e}$}\\
		\vskip 0.3in

		${}^{a}${\it Physique Math\'ematique des Interactions Fondamentales, Universit\'e Libre de Bruxelles\\ Campus Plaine 231, 1050 Bruxelles, Belgium}
        \vskip .5mm
		${}^{b}${\it Department of Physics \& Astronomy, Vanderbilt University\\ Nashville, Tennessee 37212, USA}
        \vskip .5mm
        ${}^{c}${\it Physique Th\'eorique, Universit\'e Libre de Bruxelles\\ Campus Plaine 231, 1050 Bruxelles, Belgium}
        \vskip .5mm
        ${}^{d}${\it Service de Physique de l'Univers, Champs et Gravitation, Universit\'e de Mons \\ Facult\'e des Sciences, 20, Place du Parc, B-7000 Mons, Belgium}
        \vskip .5mm
        ${}^{e}${\it Department of Physics and Beyond: Center for Fundamental Concepts in Science\\
        Arizona State University, Tempe, Arizona 85287, USA}
        \vskip 0.3in

		\texttt{sdetourn@ulb.ac.be, sahaja.kanuri@ulb.be, alexandru.v.lupsasca@vanderbilt.edu, philippe.spindel@ulb.be, quentin.vandermiers@ulb.be, rwutte@hep.itp.tuwien.ac.at}
	\end{center}
			
	\vskip 0.2in
			
	\begin{center} {\bf ABSTRACT} \end{center}
	Light circling around an astrophysical black hole can spend a long time skirting its unstably bound photon orbits before escaping to infinity.
    To a distant observer, this orbiting light would appear as a bright ring encircling the image of the black hole.
    Though not yet resolved by radio-interferometric observations from the ground, this ``photon ring'' will be the target of future space-based black hole observations.
    Motivated by this experimental prospect, studies have sought to elucidate the theoretical connections between the photon ring---an observable, classical effect---and the putative holographic description of black holes in quantum gravity.
    General relativity predicts that the detailed structure of the photon ring encodes the high-frequency (eikonal) spectrum of quasinormal modes (QNMs) emitted by a perturbed black hole as it rings down, and also that the photon ring displays an emergent conformal symmetry that acts upon this spectrum.
    In holography, the classical QNM frequencies are expected to map to Ruelle resonances of the dual quantum theory.
    In this paper, we explore these connections in a lower-dimensional toy model based on Warped AdS$_3$ black holes that shares many features with the (3+1)-dimensional Kerr background---including a photon ring at finite radius---while still providing analytic control of the QNM frequencies.
    
    \vfill\noindent\today

	\end{titlepage}

\setcounter{tocdepth}{2}
\tableofcontents

\section{Introduction}

In 2019, the Event Horizon Telescope (EHT) collaboration released the first horizon-scale images of a black hole: the supermassive black hole M87*, located 50 million light years away at the center of the galaxy Messier 87 \cite{Akiyama_2019}.
Two years later, the EHT published the first images of the supermassive black hole Sagittarius A* (Sgr\,A*), located 26,000 light years away at the center of our own galaxy, the Milky Way \cite{EventHorizonTelescopeCollaboration_2022}.
Following these breakthroughs, there has been a growing effort to improve the resolution of these images by extending the EHT ground array to space.

The Black Hole Explorer (BHEX) is a proposed space mission to launch a satellite into medium Earth orbit, at $\gtrsim20\,000\,$km altitude \cite{Johnson:2024ttr,Lupsasca:2024xhq}.
By leveraging such a distant space element and increasing the observing frequency, future radio observations using very-long-baseline interferometry (VLBI) will image these two black holes with a fivefold improvement in resolution, producing the sharpest images in astronomy \cite{Johnson:2024ttr}.
BHEX will also image a dozen previously unseen black holes.

A primary target of BHEX observations will be the measurement of the ``photon rings'' around M87* and Sgr\,A* \cite{Lupsasca:2024xhq}.
These narrow ring-shaped features---predicted by general relativity \cite{Gralla:2019xty,Johnson:2019ljv} but not yet observed \cite{Lockhart:2022rui,Tiede:2022grp}---arise from photons that orbited the black hole (possibly multiple times) before escaping to a distant observer, carrying an imprint of the black hole's strong gravity.

This experimental prospect naturally raises a question for black hole theorists:
\begin{center}
    ($\star$) \emph{What is the link between the photon ring of a black hole and its putative holographic dual?}
\end{center}

At the same time, gravitational-wave observations with LIGO \cite{LIGOScientific:2016aoc} and Virgo \cite{VIRGO:2012dcp} have measured the ``ringdown'' phase following a black hole merger, in which the waveform decomposes into a sum of exponentially decaying quasinormal modes (QNMs).
Each QNM is characterized by its complex frequency, whose real part determines the wavelength of oscillations and whose imaginary part controls their damping rate.
QNMs are intrinsically linked to the mass and spin of a black hole, which makes them vital for measuring these parameters (for reviews of these relations, see \cite{Kokkotas:1999bd,Nollert:1999ji,Berti:2009kk,Berti:2025hly} and references within).
Consequently, the ``black hole spectroscopy'' program proposes to extract QNM frequencies from gravitational-wave data in order to provide precise information about a black hole’s characteristics and test the predictions of general relativity in the strong-field regime.

In the ``eikonal limit'' where the frequency and angular momentum of QNMs grow very large, their wavefronts can be approximated by null geodesic congruences.
Since the QNMs are the last perturbations to leave the black hole, their corresponding null geodesics must orbit around it before escaping to infinity.
As a result, the spectrum of eikonal QNMs is approximated using properties of the unstable photon orbits that make up the photon ring.
Specifically, the real part of the eikonal QNM frequencies is set by the orbital frequencies of photon bound around the black hole, while the imaginary part is controlled by the Lyapunov exponents governing the instability of these unstably bound orbits.
This link between the photon ring and the eikonal QNM spectrum provides insights into the relationship between perturbations of a black hole and the behavior of light in its vicinity.

Moreover, it has been suggested that this relationship could shed some light on fundamental questions at the theoretical frontier and form a bridge between quantum-theoretical and observational black hole physics.
In the past, the quest to elucidate some of the most intriguing properties of black holes has led to the formulation of the holographic principle \cite{tHooft:1993dmi,Susskind:1994vu,Maldacena:1997re}, from which we now derive our best understanding of quantum gravity.
In that context, it has been argued that the photon ring could be a part of the holographic dual for an astrophysical black hole, encoding Ruelle resonances of the quantum dual \cite{Hadar:2022xag} and offering a partial answer to the question ($\star$).
However, this proposal has been mostly explored for black holes in Anti-de Sitter (AdS) spacetimes \cite{Kapec:2022dvc,Dodelson:2020lal,Dodelson:2022eiz,Dodelson:2023nnr}.

Directly addressing question ($\star$) for Kerr black holes---even the rapidly spinning ones---is hard.
For one, the nature of the Kerr holographic dual (if it exists), though several candidates have been proposed \cite{Guica:2008mu,El-Showk:2011euy,Guica:2017lia,Detournay:2012pc}).
More pragmatically, the full QNM spectrum of Kerr (or even Schwarzschild) is not known analytically and one has to resort to a numerical analysis \cite{Berti:2009kk, Konoplya:2011qq} (though approximate formulas exist in various limits other than the eikonal one \cite{Motl:2002hd,Motl:2003cd,Kapec:2024lnr}).
A simpler, analytically tractable toy model would therefore be highly desirable.
Spacetimes with conformal symmetry provide good candidates for such toy models, especially in the simpler lower-dimensional setting.

Lower-dimensional spacetimes have for many years proved helpful in getting a handle on some of the deepest puzzles in black hole physics. 
The three-dimensional BTZ black hole \cite{Banados:1992wn,Banados:1992gq} has provided myriad insights into the inner workings of holography and the celebrated AdS/CFT correspondence \cite{Maldacena:1997re,Gubser:1998bc,Witten:1998qj}.
Important milestones include the observation that the BTZ black hole entropy can be reproduced by counting the microstates of a hot fluid of interacting particles in a two-dimensional conformal field theory (CFT$_2$) \cite{Strominger:1997eq,Hartman:2014oaa}, and the quantitative matching of the BTZ QNM frequencies to the characteristic (Ruelle) resonances of the dual CFT$_2$ thermal state \cite{Birmingham:2001pj,Birmingham:2001hc}.

In the archetypical BTZ black hole toy model, the exact QNM spectrum is known analytically \cite{Birmingham:2001pj,Birmingham:2001hc,Dias:2019ery}, but the photon ring is pushed out to infinite radius and its physical role is therefore unclear.
A certain deformation of the BTZ black hole known as the warped AdS$_3$ (WAdS$_3$) black hole \cite{Anninos:2008fx,Moussa:2003fc,Nutku:1993eb,Gurses:1994bjn,Detournay:2005fz} provides an interesting compromise.
As we shall see, their geometry possesses a photon ring at finite radius, while still being simple enough that the full QNM spectrum is analytically computable.
However, there is a price to pay: the deformation destroys the AdS asymptotics \cite{Jugeau:2010nq,Bieliavsky:2024hus} and the holographic interpretation becomes less clear; see, e.g., \cite{Detournay:2012pc,Compere:2014bia,Apolo:2018qpq} for various (possibly interrelated) proposals.
The WAdS$_3$ black hole will be the main focus of this paper.
Recent work \cite{Kapec:2022dvc} carried out a similar analysis in the context of a related toy model, the self-dual WAdS$_3$ spacetime, which arises as the near-horizon (near-)extreme geometry of the WAdS$_3$ black hole.
Here, we extend the results of \cite{Kapec:2022dvc} to the full WAdS$_3$ black hole and show how they may be recovered in a suitable limit.

This paper takes another step towards answering the question ($\star$) in the more tractable context of the Warped AdS/CFT correspondence by working out the details of the gravity side.
Presently, the dual CFT computation is still missing, but it could in principle be carried out independently and will be the subject of future work.
Completing this program would provide an explicit example of holography in which one could definitively conclude that the photon ring is part of the hologram.

\subsection{Summary} 

This paper is organized as follows.
In Sec.~\ref{sec:PhotonRingPrimer}, we review the Kerr photon ring and its connection to QNMs, covering in detail many of the ideas mentioned above.
We pay special attention to the extreme Kerr black hole and its near-horizon limit, in which the self-dual WAdS$_3$ geometry appears.

Extreme Kerr shares many features with the WAdS$_3$ black hole, which is the subject of Sec.~\ref{sec:WBTZ}.
We calculate the location of its two photon rings in Sec.~\ref{subsec:PhotonRing}, where we also define a ``near-ring limit'' that zooms into the region of phase space near the photon ring. 
In Sec.~\ref{subsec:WarpedQNMs}, we solve the massless scalar wave equation on the WAdS$_3$ background and determine its exact spectrum of QNMs, a problem previously studied in \cite{Chen:2009rf,Martin:2022duk}.
To define QNMs, we impose the usual ingoing boundary conditions at the horizon, but we are led to consider different boundary conditions at infinity (either finite-flux, outgoing, or Dirichlet) that would normally coincide in flat space.
We relate the resulting modes and their frequencies to the parameters of the photon ring in the eikonal limit.
While finite-flux and Dirichlet conditions define the same QNMs, the outgoing-waves condition leads to a different set of resonant modes.
Intriguingly, in the eikonal limit, we find that the first spectrum is governed by the inner photon ring, while the second is controlled by the outer one.

In the remaining sections, we examine different ways to derive the eikonal QNM spectrum and its underlying symmetries.
In Sec.~\ref{subsec:PenroseLimit}, we explore the Penrose limit of the geometry near the photon ring and determine the QNM spectrum of the resulting spacetime.
Then in Sec.~\ref{subsec:EikonalSymmetry}, we analyze the QNM spectrum within the near-ring region of phase space.
Next, Sec.~\ref{subsec:GeometricOptics} uses the geometric-optics limit to approximate high-frequency solutions to the massless scalar wave equation using null geodesic congruences, before rederiving the QNM spectrum.
As expected, all these approximations coincide with each other and with the eikonal limit of the exact spectrum.

In Sec.~\ref{subsec:ConformalSymmetry}, we turn our attention away from QNMs to the conformal symmetry of the photon ring (in both phase space and in the black hole image).
In Sec.~\ref{subsec:ExtremalLimits}, we compute the QNM spectrum of the extreme and near-extreme WAdS$_3$ black holes in the near-horizon limit, in which the geometry reduces to self-dual warped AdS$_3$.
Our results are consistent with those of \cite{Kapec:2022dvc}, which we recover in the appropriate limit.
Finally, App.~\ref{app:QuadraticEnsemble} calculates the QNM spectrum for the WAdS$_3$ black hole in a different set of coordinates (the so-called ``quadratic ensemble''), which enables a comparison of our results to the BTZ QNM spectrum \cite{Birmingham:2001pj} in the limit where the warping disappears.

\section{A primer on the Kerr photon ring}
\label{sec:PhotonRingPrimer}

In this section, we review key properties of black hole photon rings and their connection to QNMs.
We pay special attention to the extreme Kerr black hole and its near-horizon geometry, which shares many features with the warped AdS$_3$ backgrounds that will be the object of Sec.~\ref{sec:WBTZ}. 
Our discussion is primarily classical, but also mentions some speculative quantum aspects near the end.

\subsection{Kerr photon shell}

General relativity predicts that astrophysical (Kerr) black holes are surrounded by a ``photon shell'': a region of spacetime that lies outside of the event horizon but in which gravity is nonetheless so strong that light can become trapped on (unstably) bound photon orbits \cite{Bardeen:1973tla,Teo:2003ltt}.
In Boyer-Lindquist coordinates with $\Delta(r)=r^2-2Mr+a^2$ and $\Sigma(r,\theta)=r^2+a^2\cos^2{\theta}$, the Kerr line element is
\begin{align}
    \label{eq:Kerr}
    ds^2=-\frac{\Delta}{\Sigma}\left(\ed t-a\sin^2{\theta}\ed\phi\right)^2+\frac{\Sigma}{\Delta}\ed r^2+\Sigma\ed\theta^2+\frac{\sin^2{\theta}}{\Sigma}\left[\left(r^2+a^2\right)\ed\phi-a\ed t\right]^2,
\end{align}
with $a$ the black hole spin. In these coordinates, the bound photons describe orbits of fixed radius.

For a nonrotating Schwarzschild black hole of mass $M$, the event horizon has radius $r_+=2M$ and photons can only orbit on a ``photon sphere'' of radius $\tilde{r}_0=3M$.
As one considers rotating black holes of increasing angular momentum $J=aM=a_*M^2$, the radius $r_+=M+\sqrt{M^2-a^2}$ of the event horizon shrinks, while the photon sphere ``thickens'' into a shell $\tilde{r}_-\le r\le\tilde{r}_+$ with 
\begin{align}
    \label{eq:ShellEdges}
    \tilde{r}_\pm=2M\left[1+\cos\left(\frac{2}{3}\arccos\left(\pm a_*\right)\right)\right].
\end{align}
At the maximal allowed spin $a=M$ (corresponding to an extreme black hole that saturates the Kerr bound $|J|\le M^2$), the shell extends from $\tilde{r}_+=4M$ down to $\tilde{r}_-=M$, seemingly touching the horizon at $r_+=M$.
However, this is deceptive, as the innermost photon orbit lies at a finite proper distance from the event horizon \cite{Bardeen:1972fi}---this extremal case will be reexamined more carefully below.

\begin{figure}
    \centering
    \includegraphics[width=0.33\textwidth]{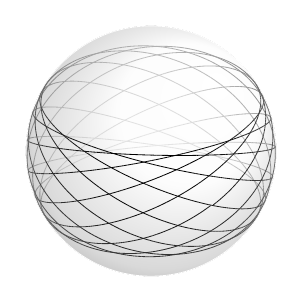}
    \vspace{-10pt}
    \includegraphics[width=0.33\textwidth]{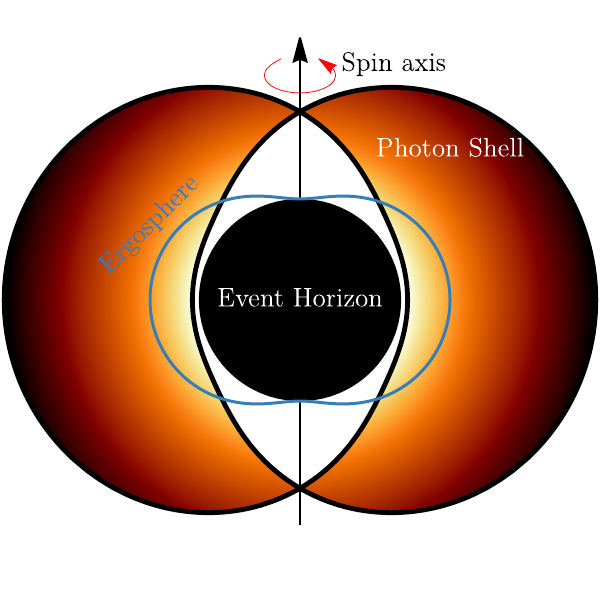}\\    \includegraphics[width=.65\textwidth]{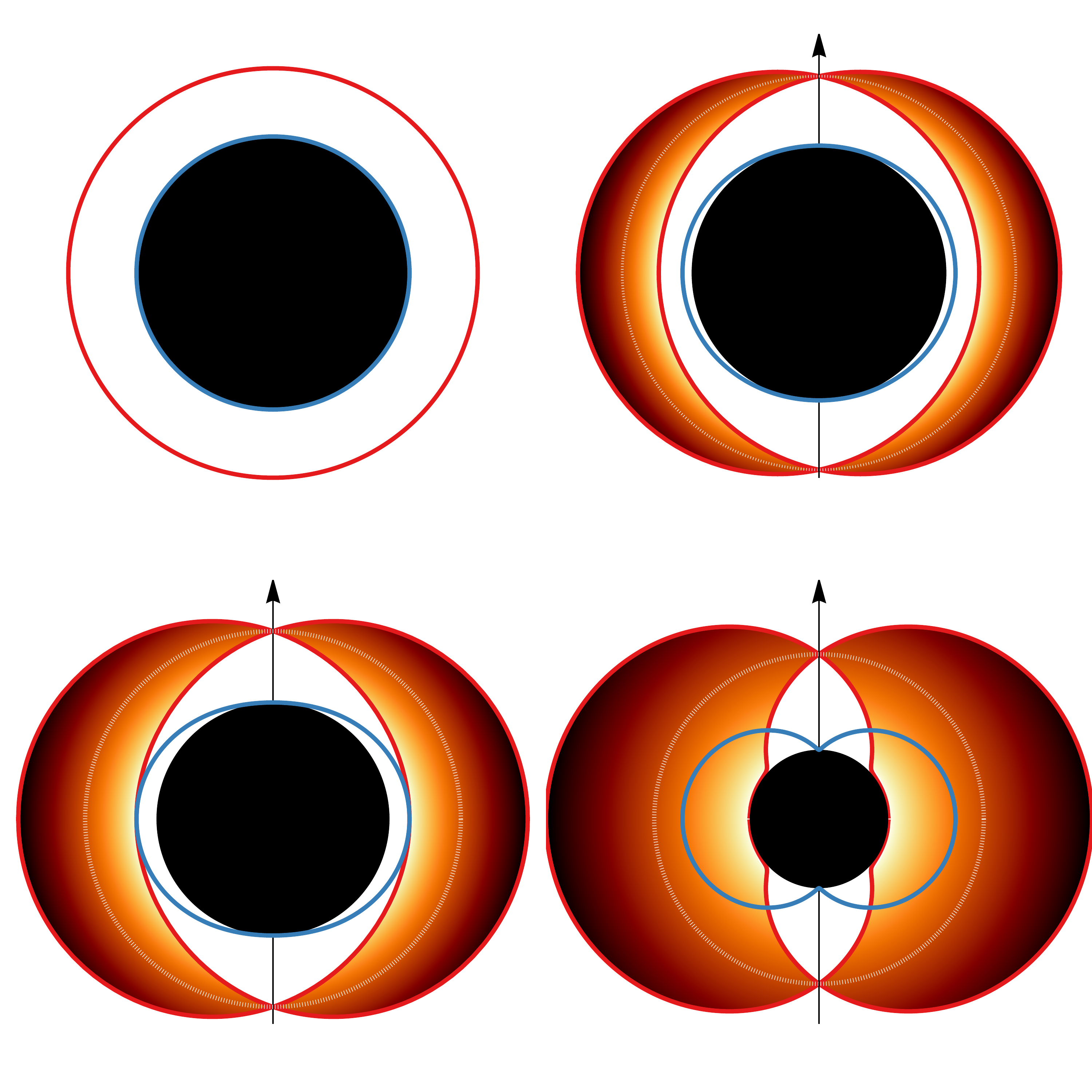}
    \caption{Top left (reproduced from \cite{Teo:2003ltt}): A bound orbit in the photon shell. This panel illustrates a segment of a generic orbit in Boyer-Lindquist coordinates; aside from a measure-zero set of closed orbits, generic orbits remain at fixed radius and oscillate between polar turning points \eqref{eq:TurningPoints}, densely filling out their shell.
    Top right (adapted from Fig.~2 of \cite{Johnson:2019ljv}): The photon shell occupies a region outside the event horizon, which we show here projected to $(r,\theta)$ surfaces for spin $a_*=94\%$.
    Middle left: For a non-rotating (Schwarzschild) black hole, the bound orbits fill out a photon sphere at $\tilde{r}_0=3M$.
    Middle right: As the black hole spin increases to $a_*=50\%$, the photon sphere thickens into a shell of radial range \eqref{eq:ShellEdges}.
    Bottom left: When the spin reaches $a_*=\frac{1}{\sqrt{2}}\approx71\%$, the innermost shell touches the ergosphere (depicted in blue).
    Bottom right: As the spin approaches extremality $a_*\to1$, the range of the photon shell extends to $[M,4M]$, appearing to touch the horizon; in reality, this never happens, though $\approx27\%$ of the photon shell [see Eq.~\eqref{eq:ExtremalPhotonShell}] fills the throat---see Fig.~\ref{fig:Throat}.}
    \label{fig:PhotonShell}
\end{figure}

Besides its energy $E=-p_t$, a Kerr photon possesses two other conserved quantities: its angular momentum about the spin axis $L=p_\phi$, and its Carter constant $Q=p_\theta^2-a^2E^2\cos^2{\theta}+L^2\cot^2{\theta}$. 
A photon that reaches a radius $r\in[\tilde{r}_-,\tilde{r}_+]$ is trapped there if (and only if) its specific spin angular momentum $\lambda=L/E$ and specific Carter constant $\eta=Q/E^2$ take the ``critical'' values
\begin{align}
    \label{eq:CriticalParameters}
    \tilde{\lambda}=a+\frac{r}{a}\left[r-\frac{2\Delta}{r-M}\right],\quad
    \tilde{\eta}=\frac{r^{3}}{a^2}\left[\frac{4M\Delta}{(r-M)^2}-r\right],\quad
    \Delta=r^2-2Mr+a^2.
\end{align}
For this reason, it is more appropriate to regard the photon shell as a region of \emph{phase space} rather than of \emph{spacetime} \cite{Hadar:2022xag}.
Recent works \cite{Raffaelli:2021gzh,Hadar:2022xag,Kapec:2022dvc} have established that the photon shell displays an emergent conformal symmetry in phase space.
The emergence of conformal symmetry is a hallmark of critical behavior and part of the modern definition of critical phenomena.
Tuning the parameters $(\lambda,\eta)$ of a photon to the ``critical'' values \eqref{eq:CriticalParameters} pushes it closer to the photon shell in phase space, dialing it into this critical locus.

Only bound photons with zero angular momentum $\tilde{\lambda}=0$ are allowed to pass over the poles. 
This can happen only at the zero-angular-mometum orbital radius $\tilde{r}_0\in[\tilde{r}_-,\tilde{r}_+]$ given by
\begin{align}
    \label{eq:ZeroAngularMomentumOrbit}
    \tilde{r}_0=M+2M\sqrt{1-\frac{a_*^2}{3}}\cos\left[\frac{1}{3}\arccos{\frac{\left(1-a_*^2\right)}{\left(1-a_*^2/3\right)^{3/2}}}\right].
\end{align}
All other bound photon orbits are limited to exploring a range of polar angles $\tilde{\theta}_-\le\theta\le\tilde{\theta}_+$, where
\begin{align}
    \label{eq:TurningPoints}
    \tilde{\theta}_\pm=\arccos\left(\mp\sqrt{\tilde{u}_+}\right),\quad
    \tilde{u}_\pm=\tilde{\triangle}\pm\sqrt{\tilde{\triangle}^2+\frac{\tilde{\eta}}{a^2}},\quad
    \tilde{\triangle}=\frac{1}{2}\left(1-\frac{\tilde{\eta}+\tilde{\lambda}^2}{a^2}\right).
\end{align}
The orbits in the innermost part of the shell $\tilde{r}_-\le r<\tilde{r}_0$ have $\tilde{\lambda}>0$ and are \emph{prograde} (i.e., they co-rotate with the black hole).
Conversely, the orbits in the outermost part of the shell $\tilde{r}_0<r\le\tilde{r}_+$ have $\tilde{\lambda}<0$ and are \emph{retrograde} (i.e., they counter-rotate relative to the black hole).

Excluding a measure-zero set of orbital radii, the period of polar librations (oscillations) in the $\theta$ direction is generically not a rational multiple of the period of azimuthal winding in the $\phi$ direction.
As a result, almost every bound photon explores the entirety of its orbital shell (though there is a measure-zero set of closed orbits), so we may identify its trajectory with the entire shell.

To summarize, bound photon orbits fill the spacetime region $r\in[\tilde{r}_-,\tilde{r}_+]$, $\theta\in[\tilde{\theta}_-(r),\tilde{\theta}_+(r)]$, $\phi\in[0,2\pi)$ for all times $t\in(-\infty,+\infty)$, as illustrated in Fig.~\ref{fig:PhotonShell}.
This region of spacetime and the momentum conditions $\lambda=\tilde{\lambda}$ and $\eta=\tilde{\eta}$ define the photon shell as a locus in geodesic phase space.

\subsection{Kerr critical curve}
\label{subsec:CriticalCurve}

\begin{figure}
    \centering
    \includegraphics[width=\textwidth]{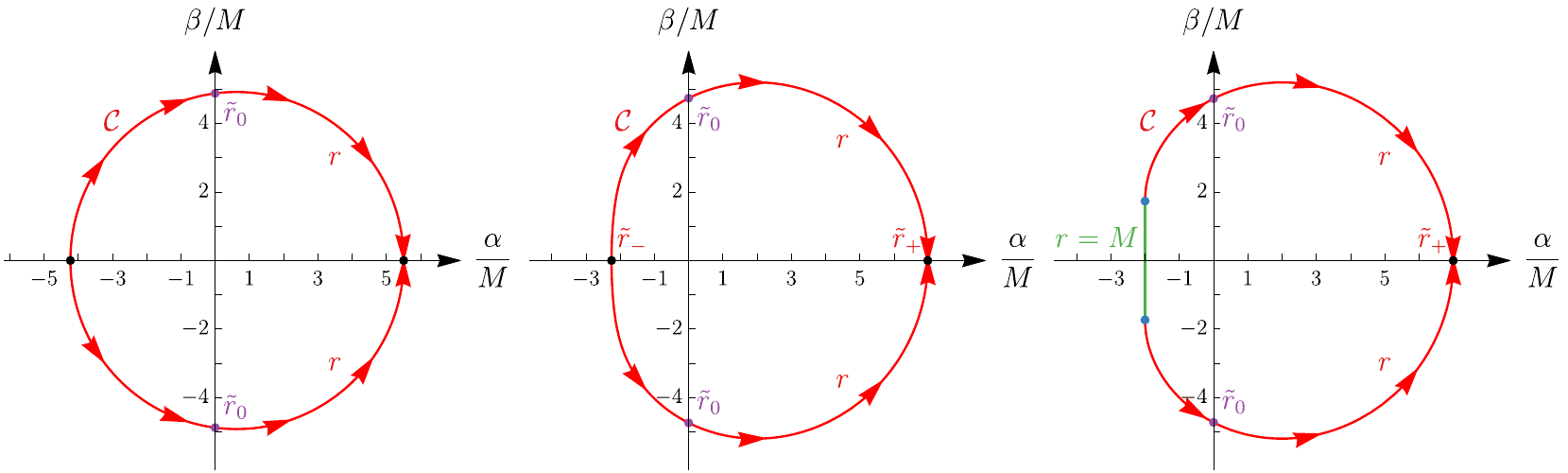}
    \caption{Image-plane critical curves for different black hole spins $a_*$ and observer inclinations $\theta_{\rm o}$: $a_*=94\%$ and $\theta_{\rm o}=17^\circ$ (left), $a_*=99.9\%$ and $\theta_{\rm o}=90^\circ$ (middle), and $a_*=100\%$ and $\theta_{\rm o}=90^\circ$ (right).
    Unless the observer lies at high inclination and the black hole has high spin, the critical curve is a nearly circular ellipse \cite{Gralla:2020yvo}.
    It is parameterized by photon shell radius $r$ and intersects the projected spin axis when $r=\tilde{r}_0$ (purple).
    Only equatorial observers see the full shell $[\tilde{r}_-,\tilde{r}_+]$ (while non-equatorial observers only see the subshell of radii such that $\tilde{\eta}\ge0$ \cite{Johnson:2019ljv}).
    At extremality (right), the critical curve develops a vertical edge (green).
    This ``NHEKline'' \cite{Gralla:2017ufe} is the image of photons that are asymptotically bound on orbits in the extremal throat (see Fig.~\ref{fig:Throat} below).
    Points on the critical curve usually correspond to double roots of the radial geodesic potential, but the edges of the NHEKline (blue) are associated with triple roots \cite{Gralla:2019ceu}.
    As the equatorial extremal observer moves to lower inclinations, the NHEKline shrinks and eventually closes at $\theta_{\rm o}=\theta_{\rm c}\equiv\arctan\left[(4/3)^{1/4}\right]\approx47^\circ$, where the spacetime geometry of the throat \eqref{eq:NHEK} reduces precisely to that of (unwarped) AdS$_3$.}
    \label{fig:CriticalCurve}
\end{figure}

In 1973, Bardeen \cite{Bardeen:1973tla} investigated the appearance of a Kerr black hole seen by a distant observer---soon after, Luminet \cite{Luminet:1979nyg} ray traced the first image of a black hole accretion disk (see also \cite{Marck:1995kd} for an early simulated \textit{movie}).
Bardeen introduced image-plane Cartesian coordinates $(\alpha,\beta)$ defined such that a photon reaching an observer at inclination $\theta_{\rm o}$ with momentum $p_\mu$ appears at image position
\begin{align}
    \label{eq:ImagePlane}
    \alpha=-\frac{\lambda}{\sin{\theta_{\rm o}}},\quad
    \beta=\pm_{\rm o}\sqrt{\eta^2+a^2\cos^2{\theta_{\rm o}}-\lambda^2\cot^2{\theta_{\rm o}}},\quad
    \pm_{\rm o}=\sign\left(p_\theta\right).
\end{align}
The origin of this coordinate system can be viewed as a ``line of sight to the center of the black hole'' and the vertical $\beta$-axis as the projection of the black hole spin axis onto the plane perpendicular to this line of sight (see App.~E of \cite{Gralla:2017ufe} for an illustration).
Bardeen also derived an analytic expression for the ``apparent boundary'' of the black hole: the mathematical curve in the image plane of an observer corresponding to light rays that asymptote to bound photon orbits around the black hole.
This image traced by critical photons is the ``critical curve'' \cite{Gralla:2019xty,Johnson:2019ljv}.
Since asymptotically trapped photons have critical parameters $(\lambda,\eta)=(\tilde{\lambda},\tilde{\eta})$, the critical curve $\mathcal{C}=\{(\tilde{\alpha},\tilde{\beta})\}$ is the parametric curve obtained by plugging Eqs.~\eqref{eq:CriticalParameters} into Eqs.~\eqref{eq:ImagePlane}.
Tracing this parametric curve for $r\in[\tilde{r}_-,\tilde{r}_+]$ and both choices of sign $\pm_{\rm o}$ results in a closed convex curve \cite{Gralla:2020yvo}, as shown in Fig.~\ref{fig:CriticalCurve}.

\begin{figure}
    \centering
    \includegraphics[width=.45\textwidth]{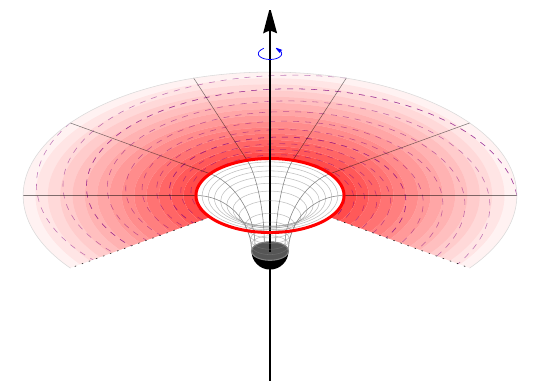}
    \includegraphics[width=.45\textwidth]{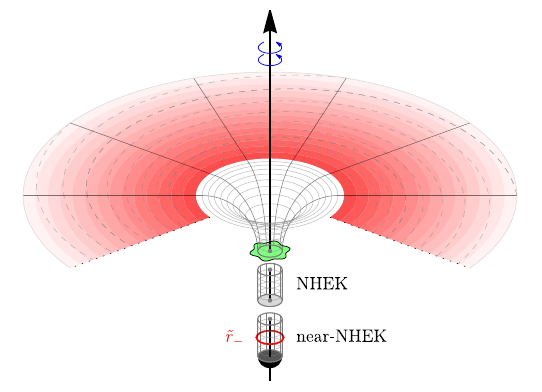}
    \caption{Embedding diagrams for the Kerr equatorial plane $\theta=\pi/2$.
    A sub-extremal ($|a|<M$) black hole has finite proper radial distance to the horizon $r=r_+$ (left), but as its spin $a=M\sqrt{1-\kappa^2}$ tends towards extremality ($\kappa\to0$), it develops a infinite throat (of log-divergent-in-$\kappa$ proper depth) that is all bunched up near Boyer-Lindquist radius $r=M$, which becomes a singular coordinate \cite{Bardeen:1972fi}.
    This singularity can be resolved via an appropriate scaling limit, resulting in the Near-Horizon Extreme Kerr (NHEK) geometry \eqref{eq:NHEK} \cite{Bardeen:1999px}.
    An extremal ($\kappa=0$) black hole has only one ``NHEK'' but a near-extremal one with $\kappa\ll1$ has multiple NHEKs atop a ``near-NHEK'' that contains both the horizon radius $r_+$ and prograde circular orbit $\tilde{r}_-$ \cite{Kapec:2019hro}.
    The extremal throat contains about 27\% 
    of the bound photon orbits, whose image forms the ``NHEKline'' (see right panel in Fig.~\ref{fig:CriticalCurve} below).}
    \label{fig:Throat}
\end{figure}

Physically, the interior of $\mathcal{C}$ corresponds to the apparent cross-section of the black hole: light rays shot back towards the black hole in those directions eventually cross its event horizon.
By contrast, the exterior of $\mathcal{C}$ corresponds to light rays that are deflected by the black hole but whose backwards extension reaches asymptotic infinity.
Hence, the critical curve is the boundary delineating the region of photon capture from that of photon escape: the ``apparent boundary'' of the black hole.

Every point on the critical curve is the image of an entire orbital shell $r\in[\tilde{r}_-,\tilde{r}_+]$, since a photon shot backwards into the geometry from a point on the critical curve asymptotes to a bound photon orbit that eventually explores the entirety of its shell.
Thus, by contrast with a star, varying the polar angle $\varphi$ in the image plane (looking around the black hole critical curve) is not equivalent to varying the azimuthal angle $\phi$ in the geometry (looking around the black hole in spacetime).
Rather, it is more akin to varying the radius in the geometry (peering closer to the black hole): this is the warped nature of spacetime staring the observer in the face!

Finally, we introduce an alternative parametrization of the photon shell and its image---the critical curve.
Rather than using orbital radius $r\in[\tilde{r}_-,\tilde{r}_+]$, one can also use the \emph{signed inclination}
\begin{align}
    \label{eq:SignedInclination}
    \mu(r)=\sign(\tilde{r}_0-r)\sin{\tilde{\theta}_\pm(r)}
    =\sign(\tilde{r}_0-r)\sqrt{1-\tilde{u}_+(r)}
    \in[-1,1].
\end{align}
The map $\mu:[\tilde{r}_-,\tilde{r}_+]\to[-1,1]$ is a monotonically decreasing bijection.
It assigns a \emph{positive} (resp. \emph{negative}) inclination $\mu=+1$ ($\mu=-1$) to the \emph{prograde} (resp. \emph{retrograde}) circular-equatorial orbit at $r=\tilde{r}_-$ (resp. $r=\tilde{r}_+$) and an inclination of $\mu=0$ to the zero-angular-momentum orbit at $r=\tilde{r}_0$.

\subsection{The (near-)extreme Kerr throat and warped \texorpdfstring{AdS$_3$}{AdS3}}
\label{subsec:ExtremeKerr}

As a Kerr black hole nears extremality ($a\to M$), the spacetime outside its event horizon is stretched into a throat-like geometry of increasing proper depth \cite{Bardeen:1972fi}.
At extremality, this throat becomes infinitely deep and its geometry is described by the Near-Horizon Extreme Kerr (NHEK) metric
\begin{align}
    \label{eq:NHEK}
    ds^2=M^2\left(1+\cos^2{\theta}\right)\left[-R^2\ed T^2+\frac{\ed R^2}{R^2}+\ed\theta^2+\Lambda^2\left(\ed\Phi+R\ed T\right)^2\right],\qquad
    \Lambda(\theta)=\frac{2\sin{\theta}}{1+\cos^2{\theta}},
\end{align}
which forms a vacuum solution of the Einstein equations in its own right \cite{Bardeen:1999px}.\footnote{This solution was first derived in 1967 by Carter \cite{cmp/1103841118} as a metric that allows for separation of the Hamilton-Jacobi equation.
It is of Petrov type D with vanishing divergence of the double principal null congruences, and corresponds to Kinnersley's Type IV.A solution \cite{Kinnersley:1969zza} with $m=0$, $a=1$, and $\ell=\frac{1}{2M^2}$ (his (4.27) is correct but his (4.28) is not):
\begin{align*}
    ds^2=&-\frac{r^2}{2M^2}\frac{x^4+6x^2-3}{\left(1+x^2\right)^3}\ed u^2-2\ed u\ed r+\frac{4xr}{1+x^2}\ed u\ed x-\frac{r}{M^2}\frac{1-x^2}{\left(1+x^2\right)^2}\ed u\ed y+2M^2\frac{1+x^2}{1-x^2}\ed x^2+\frac{1}{8M^2}\frac{1-x^2}{1+x^2}\ed y^2.
\end{align*}
Setting $u=T-\frac{1}{R}$, $r=-2M^2R\left(1+\cos^2{\theta}\right)$, $x=\cos{\theta}$, $y=8M^2(\Phi-\ln{R})$ puts this metric in the form \eqref{eq:NHEK}.}
This emergent near-horizon throat is illustrated in Fig.~\ref{fig:Throat}---see Sec.~III of \cite{Kapec:2019hro} for a more detailed discussion.

The three-dimensional metric induced on slices of constant polar angle $\theta$ in the throat is then
\begin{align}
    \label{eq:SlicedNHEK}
    ds^2=-R^2\ed T^2+\frac{\ed R^2}{R^2}+\Lambda^2\left(\ed\Phi+R\ed T\right)^2,\qquad
    \Lambda\in[0,2],
\end{align}
up to a constant rescaling.
We recognize this metric as belonging to the family of warped AdS$_3$ spacetimes; more specifically, it is the ``self-dual spacelike warped AdS$_3$'' that we will recover in Sec.~\ref{subsec:ExtremalLimits} as the near-horizon limit of the extremal warped BTZ black hole.

In particular, on the slice with polar angle
\begin{align}
    \theta_{\rm c}=\arctan\left[(4/3)^{1/4}\right]
    \approx47^\circ,
\end{align}
or $\theta=\pi-\theta_c$, the warp factor $\Lambda(\theta)$ becomes unity, $\Lambda(\theta_{\rm c})=1$, and the geometry unwarps.
That is, the polar slice $\theta=\theta_{\rm c}$ of the extremal throat has precisely the same geometry as the three-dimensional Anti-de Sitter (AdS) spacetime AdS$_3$.
AdS spacetimes have been extensively studied in the context of holography, despite our universe appearing to have positive rather than negative curvature.
Nevertheless, the AdS$_3$ geometry does appear to be realized in our universe---not globally, but deep in the throat of a rapidly spinning black hole!

Before discussing the observational appearance of orbits in this geometry, it is worth mentioning that (warped) AdS$_3$ sections of the four-dimensional Kerr spacetime are present even when the black hole is not strictly extremal.
A Kerr black hole with spin $a=M\sqrt{1-\kappa^2}$ has Hawking temperature
\begin{align}
    T_{\rm H}=\frac{1}{4\pi r_+}\left(\frac{r_+-r_-}{r_++r_-}\right)
    =\frac{\kappa}{4\pi M}+\mathcal{O}\left(\kappa^2\right).
\end{align}
Thus, a precisely extremal black hole with $\kappa=0$ has an exactly vanishing temperature $T_{\rm H}=0$, and a truly infinite throat.
By contrast, a near-extreme black hole with $0<\kappa\ll1$ has a small temperature $T_{\rm H}\sim\kappa$, and its throat has a log-divergent proper depth $D\sim M|\ln{\kappa}|$.
Hence, reaching absolute zero temperature requires an infinite stretching of spacetime, which is likely unphysical.

Nevertheless, the NHEK geometry \eqref{eq:NHEK} is still of physical relevance for (realistic) near-extreme black holes.
As explained in Sec.~III of \cite{Kapec:2019hro}, such holes have a deep (log-divergent) throat whose deepest part is described by the Near-Extreme Near-Horizon Extreme Kerr (near-NHEK) metric
\begin{align}
    \label{eq:NearNHEK}
    ds^2=M^2\left(1+\cos^2{\theta}\right)\left[-\left(R^2-\kappa^2\right)\ed T^2+\frac{\ed R^2}{R^2-\kappa^2}+\ed\theta^2+\Lambda^2\left(\ed\Phi+R\ed T\right)^2\right]\,,
\end{align}
which also forms a vacuum solution of the Einstein equations in its own right.
This near-NHEK metric \eqref{eq:NearNHEK} is related to the NHEK metric \eqref{eq:NHEK} by a coordinate transformation \cite{Bredberg:2009pv}.
Its slices of constant $\theta$ have an induced metric (up to a shift $R\to R+\kappa$) that we will rediscover in Eq.~\eqref{eq:SelfDual}.
The part of the throat in the asymptotic region that is far from the horizon ($R\gg\kappa$) but also deep in the throat ($0<r-r_+\ll1$) is still described by the NHEK geometry \eqref{eq:NHEK}, as shown in the right panel of Fig.~\ref{fig:Throat}.
This is the physical context in which (warped) AdS$_3$ may be realized in nature.

\subsection{The extreme Kerr photon shell has infinite volume}

As a Kerr black hole spins up, its photon shell thickens from a single photon sphere at $\tilde{r}_0=3M$ in Schwarzschild ($a=0$) to a thick shell stretching from $\tilde{r}_-=M$ to $\tilde{r}_+=4M$ at extremality ($a=M$).
It is then natural to ask: how much of the photon shell lies within the emergent throat geometry?

The location of the innermost photon orbit (the closed, prograde, circular-equatorial orbit at $r=\tilde{r}_-$, which forms the inner edge of the photon shell) is subtle in the (near-)extremal regime.
For a precisely extremal black hole ($a=M$), the Boyer-Lindquist radii of several special orbits---including the photon orbit at $\tilde{r}_-=M$ or even the innermost stable circular-equatorial orbit (ISCO) at $\tilde{r}_{\rm ms}=M$---exactly coincide with that of the horizon at $r_+=M$ \cite{Bardeen:1972fi}.
However, these orbits cannot physically coincide (for one, because the ISCO is a timelike orbit, while the horizon is ruled by null geodesics) so this apparent convergence must be an artefact of the coordinate system.

We stress that this effect is not due to a coordinate singularity at the horizon, as it persists in other coordinate systems (e.g., Kerr-Schild) that remain regular across the horizon.
Instead, this ``singularity'' arises because the extremal spacetime decouples into two regions---the asymptotically flat part and the near-horizon throat---that cannot be simultaneously resolved by a single coordinate system \cite{Gralla:2016jfc}.
From the perspective of the asymptotically flat region (described by the extreme Kerr metric with $a=M$), the entire throat lies at $r=r_+$, and a scaling limit into the NHEK metric \eqref{eq:NHEK} is needed to resolve near-horizon physics.
To better understand this phenomenon, we consider the extremal limit $\kappa\to0$ of a slightly near-extreme black hole with $a=M\sqrt{1-\kappa^2}$.
In this case,
\begin{align}
    r_+=M(1+\kappa),\qquad
    \tilde{r}_-=M\left[1+\frac{2}{\sqrt{3}}\kappa+\mathcal{O}\left(\kappa^2\right)\right],\qquad
    r_{\rm ms}=M\left[1+2^{1/3}\kappa^{2/3}+\mathcal{O}\left(\kappa^{4/3}\right)\right].
\end{align}
The scaling limit into the deepest part of the throat---the near-NHEK geometry \eqref{eq:NearNHEK}---ensures that all Boyer-Lindquist radii scaling as $r=M(1+\kappa R)$ end up in the near-NHEK.
In particular, this includes the horizon and the innermost photon orbit.
Since this limit ``resolves'' the throat, these two radii end up at different near-NHEK radius $R$, and indeed the proper radial distance between them retains a finite $\kappa\to0$ limit: the line element integrated along the radial direction is 
\begin{align}
    ds(r_+,\tilde{r}_-)=M\ln{\sqrt{3}}+\mathcal{O}(\kappa).
\end{align}
On the other hand, in the extremal limit $\kappa\to0$, radii that scale like $r=M(1+\kappa^pR)$ with $0<p<1$ (such as the ISCO) end up at infinite proper distance from the horizon or innermost photon orbit (which lie much deeper in the throat) and also at infinite proper distance from radii that do not scale into the throat at all.
This explains the two ``breaks'' above and below the NHEK part of the throat in the right panel of Fig.~\ref{fig:Throat} (a more detailed discussion is given in Sec.~III of \cite{Kapec:2019hro}).

The upshot of this discussion is that the proper radial distance $ds(\tilde{r}_-,\tilde{r}_+)$ between the two edges of the photon shell diverges at extremality, so the photon shell grows to infinite proper volume! 

\subsection{Extremal critical curve}

At extremality, the photon shell almost completely fills up the infinitely deep near-horizon throat, and there are infinitely many orbits within the photon shell squeezed into the same Boyer-Lindquist radius $r=M$ (as shown in Fig.~\ref{fig:Throat}, right panel). 
As a result, the orbital radius is no longer a good parametrization of the photon shell as $a\to M$.

Mathematically, plotting the critical curve in the usual way described in Sec.~\ref{subsec:CriticalCurve} does not yield a completely closed curve, but rather only part of the true critical curve.
More precisely, plotting the parametric curve obtained by plugging Eqs.~\eqref{eq:CriticalParameters} into Eqs.~\eqref{eq:ImagePlane} yields the red curve $\mathcal{C}$ shown in the right panel of Fig.~\ref{fig:CriticalCurve}, which is the image of the part of the photon shell that does not enter the throat and is resolved by the extreme Kerr metric.
To close the curve, one needs to add the image of the part of the photon shell that lies in the extremal throat, and which is resolved by the NHEK geometry.
As derived in App.~A of \cite{Gralla:2017ufe}, this can be done by taking the NHEK limit of Eqs.~\eqref{eq:CriticalParameters} before plugging into Eqs.~\eqref{eq:ImagePlane}, resulting in the vertical green segment in the right panel of Fig.~\ref{fig:CriticalCurve}.
This effectively amounts to using two coordinates to cover the entirety of the extremal critical curve: the Boyer-Lindquist radius $r$ parameterizes the part of the photon shell outside of the throat, and hence the round part of the critical curve, while the NHEK radius $R$ parameterizes the part of the photon shell inside of the throat, and hence the missing vertical line.

This vertical line is not only the image of the (infinitely large) portion of the photon shell that lies in the extremal throat, it is also where the image of any object orbiting in the throat must appear---as such, it is the image of the entire NHEK and has been dubbed the ``NHEKline'' \cite{Gralla:2017ufe}.

The edges of the NHEKline---represented by the blue dots in the right panel of Fig.~\ref{fig:CriticalCurve}---have interesting mathematical properties.
Since photon shell orbits are by definition bound, their radii must be associated with double roots of the radial potential $\mathcal{R}(r)$ that governs the radial geodesic motion via $\Delta(r)p_r=\sqrt{\mathcal{R}(r)}$ \cite{Gralla:2019ceu}.
However, if (and only if) the black hole is extremal, then it is possible for precisely one bound orbit to be a \emph{triple} root of $\mathcal{R}(r)$---see the radius $\hat{r}$ in Eq.~(100) of \cite{Gralla:2019ceu}---and this is precisely the orbit whose two images produce the NHEKline edges.
Thus, the two parts of the extremal critical curve consisting of bound orbits---double roots of $\mathcal{R}(r)$---that lie in or out of the throat are separated by a special orbit at $r=\hat{r}$ that is a triple root of the potential.

This ``triple-root'' orbit has $\hat{\lambda}=2M$ and $\hat{\eta}=3M^2$.
Photons launched from a distant observer towards the black hole with these impact parameters asymptote to this special orbit.
In the image plane, these photons appear at the location of the blue dots in Fig.~\ref{fig:CriticalCurve}: they form the edges of the NHEKline, separating it from the rest of the critical curve.
An equatorial observer sees the largest portion of the photon shell and also observes the longest possible NHEKline, stretching from $\beta=-\sqrt{3}M$ to $\beta=+\sqrt{3}M$.
As the observer rises out of the equatorial plane to smaller polar inclinations $\theta_{\rm o}$, the blue dots start to converge and the NHEKline begins to shrink, since an increasingly small subshell of the photon shell allows for asymptotically bound photons to reach such high inclinations.
Finally, at $\theta_{\rm o}=\theta_{\rm c}$, the NHEKline ``closes'' to a point as its edges coalesce.

For observers at inclinations $\theta_{\rm o}<\theta_{\rm c}$, the NHEK(line) is invisible and all points on the critical curve correspond to orbits that are asymptotically bound in the portion of the photon shell outside of the throat.
For observers at inclinations $\theta_{\rm o}>\theta_{\rm c}$, photons aimed at the edges of the NHEKline describe polar trajectories that bounce up and down between $\theta_{\rm c}$ and $\pi-\theta_{\rm c}$, that is, between the polar slices on which NHEK is exactly AdS$_3$.
Therefore, such photons explore all the spacelike warped AdS$_3$ geometries with warp factor $\Lambda(\theta_{\rm o})>1$.
In this sense, all these spacetimes can be explored by photons shot into the throat of a maximally spinning black hole. (On the other hand, the slices with squashed warp factor $\Lambda(\theta)<1$ cannot be explored by photons coming from afar).

Although the orbital radius is not a good parametrization of the extremal photon shell and critical curve (because one must use two different radial coordinates, namely $r\in[\tilde{r}_-,\tilde{r}_+]$ and $R$ with infinite range), the signed inclination $\mu(r)\in[-1,1]$ remains well-behaved in the extremal limit and does provide a single parameterization of both the photon shell and critical curve.

Indeed, the edges of the NHEKline have signed inclination
\begin{align}
    \label{eq:ExtremalPhotonShell}
     \mu(M)\equiv\sqrt{1-\tilde{u}_+(M)}
     =\sqrt{3}-1
     \approx0.73.
\end{align}
As such, we see that the extremal throat contains the part of the photon shell consisting of bound photons on prograde orbits with maximal inclination $\mu=\sin{\tilde{\theta}_\pm}\in[0.73,1]$.
Thus, we see that the extremal throat captures about 27\% of the photon shell according to this parameterization, which as we will see in Sec.~\ref{subsec:Geometrization} below is the physically relevant one for describing black hole ringdown.

\subsection{Kerr critical exponents and the photon ring}
\label{subsec:CriticalExponents}

The Kerr ``critical exponents'' are parameters associated with (near-)critical photons, that is, with photons on (nearly) bound orbits.
Though these exponents are geometric properties of bound orbits localized within the photon shell, they nevertheless control the lensing behavior of the black hole and therefore play an important role in the theory of black hole images.

To describe these quantities, we must first introduce a notion of ``half-orbit'' for orbiting photons: one orbit is defined to be a full period of the polar libration, while a half-orbit corresponds to a single ``bounce'' from $\tilde{\theta}_-$ to $\tilde{\theta}_+$ (or vice versa).
Choosing to number orbits by counting periods of the polar motion (rather than the azimuthal winding number) is a choice that is ultimately justified by the observation that sources around a black hole produce multiple relativistic images that are most conveniently labeled by this orbital number \cite{Johnson:2019ljv}---see also \cite{Lupsasca:2024wkp} for a pedagogical review.

For each orbital shell $r\in[\tilde{r}_-,\tilde{r}_+]$, one can keep track of the azimuthal angle $\Delta\phi$ swept and time lapse $\Delta t$ incurred per half-orbit.
This respectively defines the critical exponents $\delta(r)$ and $\tau(r)$ \cite{Gralla:2019drh}:
\begin{align}
    \delta(r)&=\frac{2}{\sqrt{-\tilde{u}_-}}\left[\left(\frac{r+M}{r-M}\right)K\left(\frac{\tilde{u}_+}{\tilde{u}_-}\right)+\frac{\tilde{\lambda}}{a}\Pi\left(\tilde{u}_+,\frac{\tilde{u}_+}{\tilde{u}_-}\right)\right]+2\pi\Theta(r-\tilde{r}_0),\\
    \tau(r)&=\frac{2r^2}{\sqrt{-a^2\tilde{u}_-}}\left(\frac{r+3M}{r-M}\right)K\left(\frac{\tilde{u}_+}{\tilde{u}_-}\right)+2\sqrt{-a^2\tilde{u}_-}\left[K\left(\frac{\tilde{u}_+}{\tilde{u}_-}\right)-E\left(\frac{\tilde{u}_+}{\tilde{u}_-}\right)\right],
\end{align}
where $K$, $E$, and $\Pi$ respectively denote the (complete) elliptic integrals of the first, second, and third kind, respectively, while $\Theta$ is the usual Heaviside function.

A third critical exponent $\gamma(r)$ is defined as the Lyapunov exponent governing the orbital instability of unstably bound photons.
If a critical photon bound at radius $r$ is slightly pushed off its orbital shell to a radius $r+\delta r_0$, then it will skirt this orbit for some time before eventually leaving.

According to the equation of geodesic deviation near the ring, the initial radial displacement $\delta r_0$ grows exponentially.
For a sufficiently small push $\delta r_0$, the displacement after $n$ half-orbits is \cite{Johnson:2019ljv}
\begin{align}
    \label{eq:Gamma}
    \delta r_n\approx \delta r_0e^{\gamma(r)n},\qquad
    \gamma(r)=\frac{4r}{\sqrt{-a^2\tilde{u}_-}}\sqrt{1-\frac{M\Delta}{r(r-M)^2}}K\left(\frac{\tilde{u}_+}{\tilde{u}_-}\right).
\end{align}
Although the critical exponents $\gamma$, $\delta$, and $\tau$ are defined locally in the photon shell, they play a key role in black hole lensing \cite{Gralla:2019drh}.
Because a near-critical photon can skirt its nearby bound orbit many times before escaping the photon shell, it is possible for a light source in the vicinity of a black hole to produce multiple images in the sky of a distant observer, arising from photons that execute any number $n$ of half-orbits within the photon shell on their way from source to observer.
By Eq.~\eqref{eq:Gamma}, high-$n$ photons must be exponentially close (in $n$) to criticality (i.e., to their nearby bound orbit), and hence their image-plane position must be exponentially close to the critical curve.
As a result, highly lensed images of a source must accumulate very near the critical curve, forming a bright, narrow ``photon ring'' of excess flux on top of the ``direct'' emission.

In the case of a ``face-on'' observer looking at the black hole from the spin axis, it is particularly simple to describe the observational appearance of an equatorial point source (see Fig.~8 of \cite{Gralla:2019drh}): if its $n^\text{th}$ image (produced by a photon that executed $n$ half-orbits) appears on the image plane at time $t_n$, polar angle $\varphi_n$, and perpendicular distance $d_n$ from $\mathcal{C}$, then its $(n+1)^\text{th}$ image appears at
\begin{align}
    t_{n+1}=t_n+\tau(\tilde{r}_0),\qquad
    \varphi_{n+1}=\varphi_n+\delta(\tilde{r}_0),\qquad
    d_{n+1}=e^{-\gamma(\tilde{r}_0)}d_n,
\end{align}
where $\tilde{r}_0$ denotes the radius \eqref{eq:ZeroAngularMomentumOrbit} of the bound photon orbit with zero angular momentum (which is the only orbit that can be seen from infinity, since only photons with $\lambda=0$ are allowed to reach the pole $\theta=0$---in other words, the entire critical curve of a polar observer corresponds to photons that asymptote to $r=\tilde{r}_0$).
Intuitively, a photon that completes an additional half-orbit around the black hole reaches the observer after an additional time lapse $\Delta t=\tau$, having swept an extra angle $\Delta\phi=\delta$ around the black hole and hence around the image.
Thus, despite their coordinate-dependent definitions, $\gamma$, $\delta$, and $\tau$ nonetheless describe lensing by a Kerr black hole.

To summarize, light sources around a black hole produce multiple images indexed by photon half-orbit number $n$.
These images accumulate near the critical curve where they form a ``photon ring'' that is the image of near-critical photons passing close (in phase space) to the photon shell.
The photon shell critical exponents $\gamma$, $\delta$, and $\tau$ respectively control the demagnification, rotation and time delay of successive images appearing in the photon ring, and are thus observable quantities.

\subsection{Eikonal QNMs and the photon ring}
\label{subsec:Geometrization}

The critical exponents of the photon shell not only control the substructure of the photon ring, but also the eikonal (high-frequency) part of the spectrum of Kerr quasinormal modes \cite{Yang:2012he,Hadar:2022xag}.

When a black hole is perturbed, or after it coalesces out of the merger of two black holes, it rapidly returns to equilibrium by ``ringing down,'' much like a bell after it is rung.
Given the shape and composition of a bell, it is possible to compute its characteristic ``sound'' in the form of its normal modes (``notes'').
In much the same way, since the geometry of spacetime around a black hole is known to be described by the Kerr metric, it is possible to predict the characteristic notes of its ``ringdown,'' known as its quasinormal modes (QNMs).
These modes have complex frequencies (which is why they are called ``quasi''-normal) with a negative imaginary part that makes them exponentially damped---this is the statement that the excitations of a black hole die down rapidly.

From an observational perspective, LIGO (and eventually LISA) can detect the gravitational waveform produced during a black hole merger.
Such waveforms can be divided into three stages: first, an inspiral phase of relatively slow oscillations (with a frequency set by the orbital period of the system), followed by a short merger phase in which the coalescing black holes enter the fully nonlinear regime of general relativity, and finally the ringdown phase in which the final black hole settles down to equilibrium.
The exponentially damped signal in this final stage can be decomposed into a sum over Kerr QNMs, as reviewed in \cite{Berti:2009kk,Berti:2025hly}.
These QNMs are labeled by four (half-)integers $s$, $\ell$, $m$, and $n$: the first is the spin $s$ of the field being perturbed, then the next two integers are spheroidal harmonic numbers $(\ell,m)$, subject to the usual constraints $-\ell\le m\le+\ell$ and $\ell\ge s$, and finally $n\ge0$ is an overtone number, with $n=0$ denoting the fundamental mode and $n\ge1$ its higher overtones.
For each choice of these parameters, there exists a QNM with frequency $\omega_{\ell m n}^{(s)}$.

The exact spectrum $\omega_{\ell m n}^{(s)}$ of Kerr QNMs is not analytically tractable, in the sense that the QNM frequencies do not have an elementary closed-form expression (though they do have a representation in terms of supersymmetric partition functions).
In practice, one uses Leaver's method to compute them via a numerical continued-fraction method \cite{Leaver:1985ax}.
However, in certain limits, the QNM spectrum becomes amenable to analytic treatment via a number of approximations.
In particular, it is well-known that in the high-frequency (or ``eikonal'') limit, wavefronts propagate in the same way as a collection of particles.
To leading order in this geometric-optics approximation, one can thus describe any high-frequency wave as a congruence of geodesics.
Since spin effects are subleading, gravitational and electromagnetic waves are both approximated by null geodesic congruences, and we will suppress the spin even though our focus here is on gravitational perturbations with $s=2$.

Black hole QNMs are waves of definite azimuthal angular momentum $m$ and energy $\omega_{\ell m n}$, which at high frequencies scale like $\omega\sim\ell$.
The eikonal part of the QNM spectrum is therefore obtained in the limit $\omega,\ell\to\infty$ with $\bar{\mu}\equiv m/\ell$ kept fixed.
By definition, this parameter is allowed to take $2\ell+1$ values evenly distributed in the range $-1\leq\bar{\mu}\leq1$.
In the eikonal limit, however, it becomes a continuous parameter $\bar{\mu}\in[-1,1]$.

\begin{figure}
    \centering
    \includegraphics[width=0.6\linewidth]{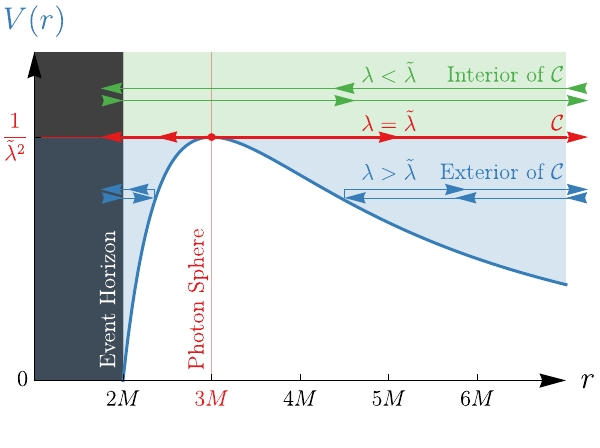}
    \caption{Null geodesics in Schwarzschild have radial trajectories obeying $\left(\frac{dr}{ds}\right)^2=\frac{1}{\lambda^2}-V(r)$ with potential $V(r)=\frac{1}{r^2}\left(1-\frac{2M}{r}\right)$.
    Only geodesic congruences with critical specific angular momentum $\tilde{\lambda}=3\sqrt{3}$ can satisfy the boundary conditions (ingoing at horizon and outgoing at infinity) required to approximate QNMs in the eikonal limit (other congruences describe modes with the wrong behavior at horizon and infinity).
    The photon ring is thus a geometrization of the QNM spectrum.}
    \label{fig:QNM}
\end{figure}

Which null congruences correspond to eikonal QNMs in the geometric optics approximation? It turns out that the QNM boundary condition picks out a very special null congruence: the set of light rays that asymptote to bound orbits in the photon shell.
The reason is the following: QNMs are defined to be purely ingoing at the horizon and purely outgoing at infinity---it is these two boundary conditions that quantize the spectrum into a discrete set of frequencies $\omega_{\ell m n}$.
From a particle perspective, null geodesics outside the critical curve describe photons that bounce off the potential barrier formed by the photon shell, so the corresponding wavefronts must have both ingoing and outgoing components (both at infinity and at the horizon).
Likewise, null geodesics inside the critical curve describe photons that can pass over the potential barrier, so the corresponding wavefronts must be either ingoing at both infinity and the horizon, or outgoing at both places.
As illustrated in Fig.~\ref{fig:QNM}, only null geodesics that asymptote to the photon shell in the far past can describe wavefronts that leak away from the potential barrier in both directions.
In this sense, the photon shell can be viewed as a geometrization of eikonal part of the QNM spectrum.

This connection between eikonal QNMs and bound photon orbits was understood in the 80s via the work of Ferrari and Mashoon \cite{Ferrari:1984zz, Mashhoon:1985cya}, who explicitly worked out this correspondence for Schwarzschild black holes.
Soon after, Iyer and Will \cite{Iyer:1986np, Iyer:1986nq} extended this WKB analysis to very high order in the eikonal expansion, but their assumed form of the geodesic potential did not include that of the Kerr black hole.
Finally, the Kerr case was only tackled in 2012 \cite{Yang:2012he}, though these authors did not provide a particularly simple expression for the eikonal QNM spectrum.

Such a formula was only very recently obtained in \cite{Hadar:2022xag} using the critical exponents from Sec.~\ref{subsec:CriticalExponents}:
\begin{align}
    \label{eq:KerrQNMs}
    \omega_{\ell m n}=\left(\ell+\frac{1}{2}\right)\Omega_{\rm R}(\bar{\mu})-i\left(n+\frac{1}{2}\right)\gamma_{\rm L}(\bar{\mu})+\mathcal{O}\left(\frac{1}{\ell}\right).
\end{align}
In this expression, we are taking the eikonal limit in which $\ell\to\infty$ (with $\omega\sim\ell$ as advertised) while $\bar{\mu}=m/\ell$ is kept fixed and becomes a continuous parameter $\bar{\mu}\in[-1,1]$.
For a given choice of $\ell$, $m$, and $n$, the QNM frequency is fixed by its real and imaginary coefficients $\Omega_{\rm R}(\bar{\mu})$ and $\gamma_{\rm L}(\bar{\mu})$, which depend only on the ratio $m/\ell$, while the overtone number controls the magnitude of the imaginary damping component (so higher overtones are exponentially more damped).
All that remains to be specified is these coefficients, and it turns out that
\begin{align}
    \Omega_{\rm R}(\bar{\mu})=\frac{\tilde{\lambda}(r(\bar{\mu}))}{\bar{\mu}},\qquad
    \gamma_{\rm L}(\bar{\mu})=\frac{\gamma(r(\bar{\mu}))}{\tau(r(\bar{\mu}))},
\end{align}
where $\tilde{\lambda}{(r)}$ is the critical specific angular momentum defined in Eq.~\eqref{eq:CriticalParameters}, $r(\mu)$ is the inverse of the bijective map $\mu(r)$ defined in Eq.~\eqref{eq:SignedInclination}, while $\gamma(r)$ and $\tau(r)$ are the Lyapunov exponent and orbital period defined in Sec.~\ref{subsec:CriticalExponents}.
This explicitly shows how the photon shell geometrizes the eikonal QNM spectrum: in this regime, the QNMs can be viewed as waves whose wavefronts are made up of photons that asymptote to bound orbits in the photon shell, with the maximal ``inclination'' $\sin{\theta}$ of the orbit determined by the ratio $m/\ell$ of the QNM.

\subsection{Top-down holography: the AdS/CFT correspondence in string theory}
\label{subsec:TopDown}

A major goal of quantum gravity is to develop a quantum-mechanical description of black holes.
As we will now review, the leading approach to this problem involves the ``holographic principle''.
Since the photon ring plays a central role in classical black hole physics, an analysis of its properties in the context of holography may provide useful clues about the quantum theory of black holes. Holography is by now a vast subject in its own right, covering too much ground to review---our central objective here is simply to clarify what is assumed versus derived, and delineate what is definitively known from what is only conjectured.
The historical development did not follow the most logical presentation, so we first discuss the ``top-down'' approach to holography and then its application to black holes, even though chronologically the latter preceded the former.
The top-down approach consists of a derivation of the holographic principle, within the specific context of AdS spacetimes and under some assumptions, starting from a candidate for a complete theory of quantum gravity: string theory.
The key insights in this approach are due to Maldacena \cite{Maldacena:1997re} and summarized in the exhaustive MAGOO review \cite{Aharony:1999ti}, which we will refer to extensively.

String theory is only well-understood as a perturbative theory, in which the fundamental object is a (closed or open) string whose vibrational degrees of freedom create all possible types of particle excitations.
At the non-perturbative level, other fundamental objects appear in the theory, most notably D$p$-branes: $p$-dimensional membranes whose time evolution produces $(p+1)$-dimensional world volumes, and that can interact with the usual strings.

We now review two key scenarios in the specific context of ten-dimensional superstring theory: the first is under complete control but not evidently related to black holes, while the second pertains to higher-dimensional black holes (branes) but lies beyond the reach of perturbative techniques.

\begin{figure}
    \centering
    \includegraphics[width=0.45\linewidth]{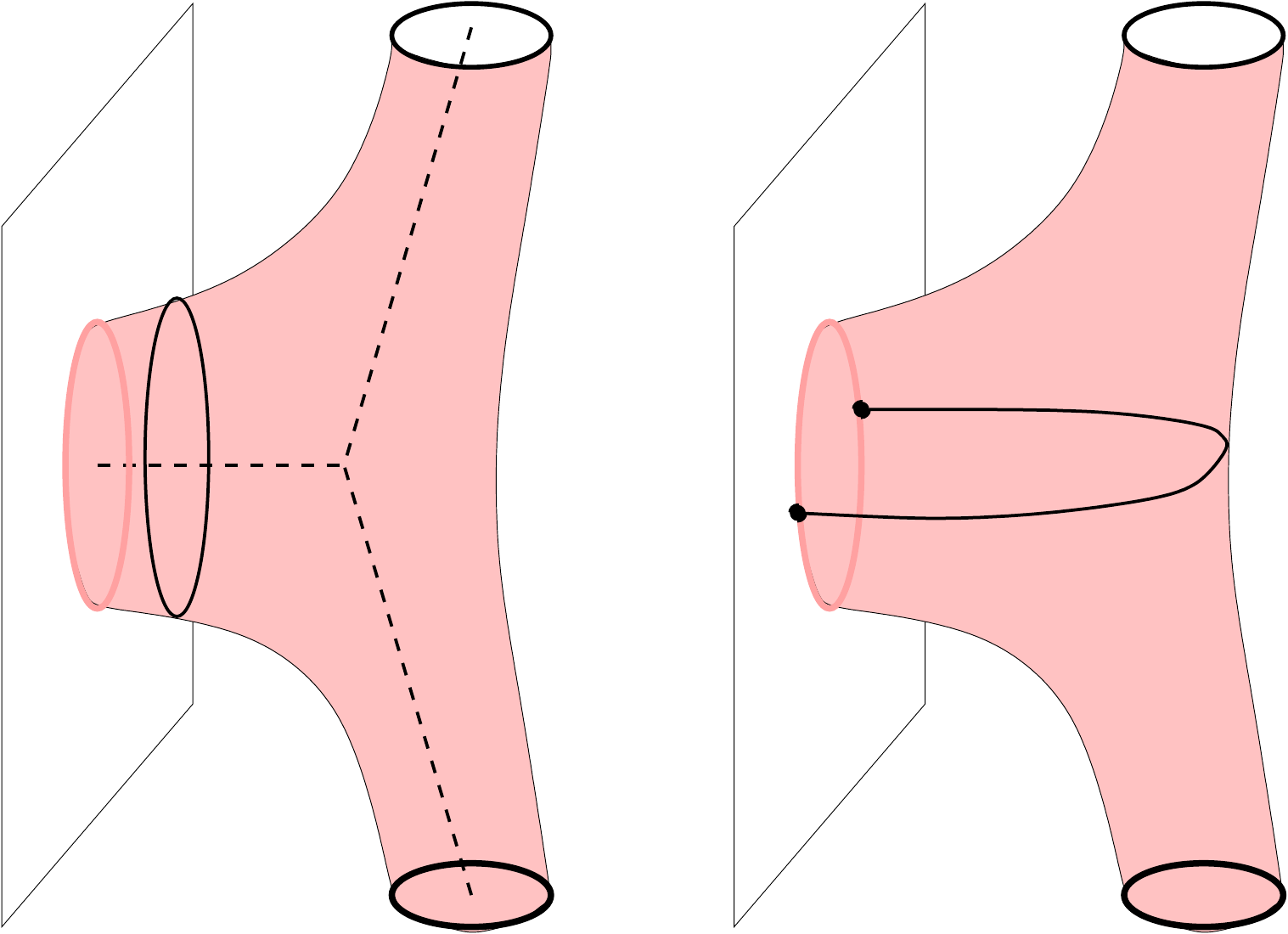}\qquad
    \includegraphics[width=0.45\linewidth]{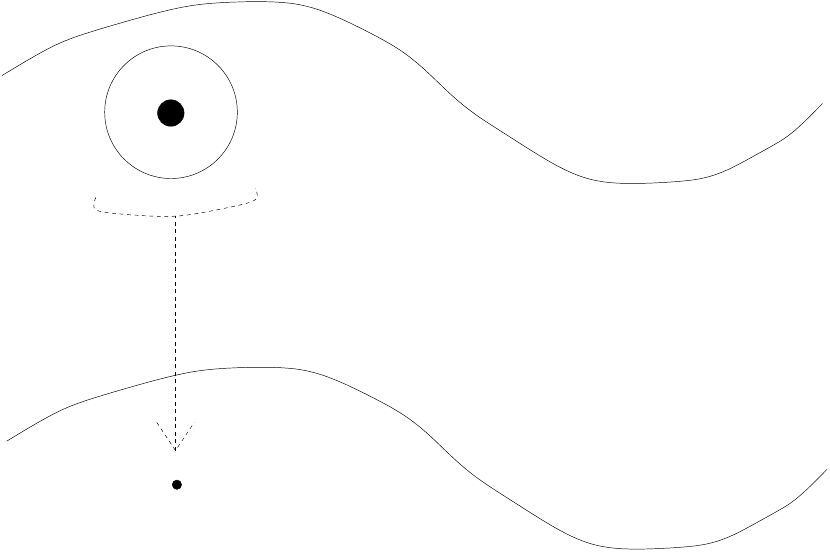}
    \caption{The left panels (reproduced from Fig.~2 of \cite{Schomerus:2006txi}) illustrate open/closed string duality.
    There are two equivalent ways to view an interaction between a string and a stack of $N$ coincident branes: either the incoming closed string is scattered by another closed string emitted by the branes, or it splits into an open string whose endpoints propagate along the stack of branes before decaying as an outgoing closed string.
    When many branes are stacked, they collapse into a ``black brane'' with an AdS throat in its vicinity.
    The right panel (reproduced from Fig.~1.4 of \cite{Aharony:1999ti}) illustrates a long-wavelength excitation scattering off the brane (top).
    In the ``decoupling limit'' $\omega\to0$ (bottom), the brane and its AdS throat shrink to a point from the perspective of the long-wavelength excitations, and physics in the throat decouples from the asymptotically flat region.}
    \label{fig:StringBraneInteraction}
\end{figure}

In the first scenario, one first considers a stack of $N$ coincident D3-branes in flat ten-dimensional Minkowski spacetime.
Gubser and Klebanov \cite{Gubser:1997se} computed the absorption cross-section $\sigma_{\rm abs}$ for a closed string interacting with these branes.
As long as the strength of the gravitational interaction, as measured by the string coupling constant $g_s$, is weak (that is, $Ng_s\ll1$), the quantum-mechanical amplitude for this process is computed by a world-sheet calculation in perturbative string theory.
To leading order in $g_s$, the tree-level amplitude is given by a two-point function of the effective theory on the D3-branes---see MAGOO Eqs.~(1.38)--(1.39) for details---which is known to be a $\mathsf{U}(N)$ gauge theory.
One interpretation is that the gravitational interaction of the closed string with the D3-branes is equivalent to a process in which the D3-branes are perturbed by the endpoints of an open string into an excited state (in the gauge theory on the branes), which then decays by emitting an outgoing closed string.
Pictorially, this ``open/closed string duality'' is illustrated by the two equivalent ways of cutting the string worldsheet in the left panels of Fig.~\ref{fig:StringBraneInteraction}: in the leftmost panel, the incoming closed string is scattered by another closed string that the D3-branes spontaneously emit, while in the next panel, the incoming closed string hits the D3-branes and splits open, then its endpoints propagate along the branes before rejoining to form an outgoing closed string.

Next, one takes the low-energy limit of this process.
In the gravitational picture, this amounts to giving the incoming string a long wavelength much larger than any characteristic curvature scale $R$ of the geometry, or equivalently, a very small frequency $\omega\ll1/R$.
In the gauge-theoretic picture, this is equivalent to letting the effective theory of the D3-branes flow deep into the infrared (IR).
Since this theory lives on the worldvolume of the D3-branes, it is a four-dimensional quantum field theory, and its IR limit must be a CFT$_4$.
One can show that this CFT$_4$ must have $\mathcal{N}=4$ supersymmetry, implying that it can be uniquely identified, already from symmetry alone, as $\mathcal{N}=4$ Super-Yang-Mills (SYM) with gauge group $\mathsf{U}(N)$ and coupling $g_{\rm YM}\sim\sqrt{g_s}$.

To summarize, one considers the gravitational interaction of a string with a stack of branes, in the double limit of weak gravity $Ng_s\ll1$ and low energy $\omega R\ll1$.
The first condition is needed to control the calculation: because $Ng_s\ll1$, this process can be completely described within perturbative world-sheet string theory, and one finds that this string-theoretic three-point function (for either scalars or gravitons, depending on the vibrational modes of the string involved) is related to a two-point function in the effective gauge theory on the branes.
The second condition is needed to take the IR limit of the effective theory of the D3-branes, which can then be exactly identified as $\mathcal{N}=4$ $\mathsf{U}(N)$ SYM.
The equivalence in this double limit between the decay of an excitation in $\mathcal{N}=4$ SYM and the stringy gravitational scattering is not a conjecture, but rather a derivable feature of perturbative string theory that holds as a consequence of open/closed string duality.

Now comes the conjectural part.
We consider instead a second scenario, in which $Ng_s\sim Ng_{\rm YM}^2$ is no longer small; that is, we increase either the strength of gravity, or the number of branes.
What happens on both sides of the duality then?
The precise answer can no longer be derived within perturbative string theory, but one can still guess it.
On the stringy side, the gravitational interactions between the branes must grow so strong that their energy backreacts on the geometry, forming a ``black brane'' (akin to a black hole with extended directions); meanwhile, on the gauge-theoretic side, the CFT$_4$ becomes strongly coupled.
In this case, the perturbative picture in Fig.~\ref{fig:StringBraneInteraction} breaks down, but one may conjecture that open/closed string duality persists at strong coupling. More precisely, when the stack of branes backreacts on the geometry, the resulting spacetime is no longer flat, but can instead be described by a particular black brane solution in supergravity, given in MAGOO Eq.~(3.3).
The near-horizon limit of this black brane, explicitly given in MAGOO Eq.~(3.5), is none other than AdS$_5\times S^5$.
In other words, the spacetime geometry near the brane ``opens up'' into an AdS throat of infinite volume, much like the near-horizon region of an extreme Kerr black hole illustrated in Fig.~\ref{fig:Throat}.
Staying in the regime of strong coupling, but taking the low-energy limit of the scattering process, physics in this throat ``decouples'' from the asymptotically flat region, as illustrated in the right panel of Fig.~\ref{fig:StringBraneInteraction}.
This observation led Maldacena to conjecture a precise form of the AdS/CFT correspondence: quantum gravity in AdS$_5\times S^5$ is dual (at weak or strong coupling) to a $\mathcal{N}=4$ SYM CFT$_4$ at the boundary of the AdS$_5$ throat.

For completeness, we mention the other precise early example of the AdS/CFT correspondence, reviewed in Sec.~5.3 of MAGOO.
In this version, we start with perturbative strings on $\mathbb{R}^5\times S^1\times\mathbb{T}^4$.
Then we add a stack of $N_1$ coincident D1-branes extending along some non-compact direction, as well as a stack of $N_5$ coincident D5-branes wrapping $\mathbb{T}^4$ and sharing the non-compact direction with the D1-branes.
At strong coupling on the gravity side, they backreact into a six-dimensional ``black string'' solution in supergravity (which upon Kaluza-Klein compactification along the longitudinal direction dimensionally reduces to a five-dimensional black hole).
This solution has the near-horizon geometry given in MAGOO Eq.~(5.14) and is none other than AdS$_3\times S^3\times M^4$.
In the low-energy limit, physics in the AdS$_3\times S^3$ throat decouples from the far region and becomes conjecturally dual to the IR limit of the effective theory of the branes: the D1-D5 CFT$_2$ living at AdS boundary.

 AdS/CFT can thus be viewed as the conjecture that the open/closed string duality derived in perturbative string theory survives at strong coupling, at least in the decoupling limit $\omega\to0$.
Because the near-horizon throat is at infinite redshift relative to the far region, any finite energy excitation in AdS still has zero energy from the perspective of the asymptotically flat region, and is therefore still captured in this limit; in other words, even though AdS/CFT is obtained from a low-energy limit of the full spacetime, it captures all the physics in AdS, including at high energies.

\subsubsection{Why is the holographic plate at the boundary?}

Naively, the dual CFT lives at the AdS boundary because that is the region of spacetime that is invariant under the conformal isometry group of the bulk spacetime.
A deeper reason is that only near the boundary does gravity attenuate so much that the background spacetime is well-defined.
If it were not for this, then the holographic plate would itself be part of the hologram.
This is one of the reasons why the Kerr/CFT conjecture (see below) is only formulated for extreme Kerr black holes with a throat and a decoupling limit---for sub-extremal black holes, it is not clear where the holographic plate should be.

Going back to the examples of an AdS throat connected to an asymptotically flat region, it is important for the existence of a decoupling limit that not much happens at the boundary where the two meet.
From the AdS perspective, this decoupling can be understood as a gravitational redshift effect: perturbations that climb out of the throat get their wavelength stretched and lose their energy near the boundary.
Thus, from the bulk perspective, the boundary lies in the IR.

\subsubsection{Deformations of AdS/CFT}

A related observation goes under the name of ``UV/IR duality'': the isometries of AdS include a dilation that pushes bulk points closer to the boundary (the bulk IR), while pushing points on the boundary to shorter distances (the ultraviolet of the boundary).
Hence, AdS/CFT maps IR to UV.

This idea points the way to deformations of AdS/CFT: modifiying the dual CFT by a relevant deformation (which preserves its UV but changes its IR) keeps the AdS boundary fixed but can change the spacetime deep in the bulk (preserving its IR but changing the bulk UV).
In that case, there is still a CFT in the UV but the deformation triggers a flow that breaks conformal symmetry.
An extreme version of this could be to replace the UV CFT by a different kind of theory, such as a warped CFT perhaps altogether, leading to a warped version of AdS/CFT relevant for Sec.~\ref{sec:WBTZ}.
Such scenarios have appeared recently in the literature involving a certain type of irrelevant deformation of a 2d CFTs \cite{Guica:2017lia, Apolo:2018qpq}.

\subsection{Bottom-up holography for black holes}
\label{subsec:BottomUp}

The AdS/CFT correspondence suggests a more general ``holographic principle'': quantum gravity in a bulk spacetime is dual to a lower-dimensional quantum field theory living on its boundary.
In hindsight, there were several discoveries within general relativity (GR) that anticipated this idea, from the Bekenstein-Hawking law stating that the entropy of a black hole scales with its area (rather than its volume), to the observation that the ADM energy (and hence the quantum Hamiltonian) of a gravitational system is defined only at its boundary.

The bottom-up approach to holography seeks to ``derive'' as many facets of the holographic correspondence as possible, starting not from a postulated UV-complete theory of quantum gravity and ``working one's way down,'' but rather ``up'' from its low-energy limit: GR.
Of course, in the absence of a microscopic theory, one should not expect such an approach to provide as many details as the explicit string-theoretic examples reviewed above.
Nevertheless, a remarkable amount can be inferred from two lines of reasoning, one based on symmetry the other on scattering experiments.
The power of symmetry is best illustrated with the original 1986 Brown-Henneaux paper \cite{Brown:1986nw}, in which they carried out---purely within GR---an asymptotic symmetry group (ASG) analysis: they imposed boundary conditions defining asymptotically AdS$_3$ spacetimes and determined the (large) diffeomorphisms preserving them.
The resulting ASG consisted of two copies of the Virasoro group.

Thus, quantizing gravity in AdS$_3$ should give a quantum theory with two Virasoro symmetries---the symmetry algebra of a CFT$_2$.
In hindsight, the Brown-Henneaux computation anticipated AdS$_3$/CFT$_2$, but it took more than a decade for its precise realization to be accepted and proposed.
Nevertheless, symmetry led to the correct answer early on.
Attempts to generalize this approach to other settings include, but are not restricted to, dS/CFT for asymptotically de Sitter spacetimes \cite{Strominger:2001pn}, the Carrollian (for a recent review on the topic and an extensive list of references, see Talk\cite{grumiller2025carroll}) and celestial holography (see, e.g.\ \cite{Raclariu:2021zjz, Donnay:2023mrd} for reviews on celestial holography)
programs for asymptotically flat spaces, and Kerr/CFT \cite{Guica:2008mu} for the extreme Kerr throat (see below).

The other bottom-up idea involves scattering experiments.
As reviewed in Sec.~1.3.3 of MAGOO, ``an important precursor to the AdS/CFT correspondence was the calculation of grey-body factors for black holes built out of D-branes.''
In the string-theoretic constructions from the last section, the grey-body factor of the black brane is encoded in the absorption cross-section $\sigma_{\rm abs}$ of a string scattering off of it, which can be expressed in terms of a CFT correlator.
This observation predates AdS/CFT, which was anticipated by black hole scattering calculations, such as the one in \cite{Maldacena:1996ix}.
The idea is to consider scattering off a six-dimensional, extreme black hole with an AdS$_3\times S^3$ throat.
The problem is reducible to a one-dimensional, Schr\"odinger-type scattering problem.
This equation is not exactly solvable, but it is amenable to analytic treatment via the method of matched asymptotic expansions, in which the spacetime is divided into a ``far'' (asymptotically flat) region and a ``near'' AdS$_3\times S^3$ throat region.
The wave equation simplifies in both regions, descending in the near region to the exact wave equation in AdS$_3\times S^3$.
By the conformal symmetry of AdS$_3$, the near wave equation is exactly soluble via hypergeometric functions, which form representations of the $\mathsf{SL}(2,\mathbb{R})$ isometry group.
An approximate solution to the full scattering problem is obtained by ``matching'' the two asymptotic ``near'' and ``far'' solutions in the region where they overlap.

It was noticed that the contribution from the near region with conformal symmetry takes the form of a correlator in a CFT$_2$.
This calculation thus suggests the following picture: one can ``cut out'' the part of the spacetime in the near-horizon throat, and replace it by a CFT living on its boundary---the ``mouth'' of the throat.
This picture is drawn in the right panel of Fig.~\ref{fig:Throat}, where the CFT can be thought of as living on the green cut.
From the perspective of the scattering problem, the interaction of low-energy waves impinging upon the black hole from afar can be equivalently described by an interaction with this CFT living on this boundary of the throat.

This picture, in which the black hole is replaced by a CFT at the mouth of its throat, in hindsight suggests a slight generalization of the AdS/CFT conjecture to small but nonzero $\omega\ll 1/R$.
In the exact limit $\omega\to0$, the near-horizon throat completely decouples from the far region, and one is left with a physical equivalence between physics in the AdS$_3$ throat and in the CFT$_2$ living on its boundary (the $S^3$ part of the geometry just comes along for the ride).
That is, one recovers the usual AdS/CFT in a decoupling limit that ``forgets'' the rest of the black hole spacetime.
But the originally discovered correspondence also applies (approximately) to leading order in small $\omega$.

Crucially, this duality includes not only the throat itself, but must also extend to a region covering a few Schwarzschild radii away from the horizon.
As first pointed out in \cite{Maldacena:1996ix}, this is in fact necessary for this conjectured holographic correspondence to hold:
\begin{quote}
The black hole emits blackbody radiation from the horizon.
Potential barriers outside the horizon act as a frequency-dependent filter, reflecting some of the radiation back into the black hole and transmitting some to infinity.
The filtering acts in just such a way that the black hole spectroscopy mimics the excitation spectrum of the string.
Hence to the observer at infinity the black hole, masquerading in its greybody cloak, looks like the string, for energies small compared to the inverse Schwarzchild radius of the black hole.
In the past, greybody factors have been largely regarded as annoying factors which mar the otherwise perfectly thermal blackbody radiation.
Now we see that they have an important place in the order of things, and transmit a carefully inscribed message on the quantum structure of black holes.
We also see that in order to compare the string and black hole pictures, we must take into account processes which occur well outside the horizon of the black hole solution.
\end{quote}
For a Kerr black hole, a region extending a few $M$ outside the horizon would also contain at least part of the photon shell,
Indeed, in the eikonal limit, these frequency-dependent potential barriers are related to the potential peaks associated with bound photon orbits, such as the one in Fig.~\ref{fig:QNM}.

\subsection{Kerr/CFT conjecture}

As we have seen, black hole analyses predating AdS/CFT suggest that the holographic principle applies not just to asymptotically AdS spacetimes, but also to the near-horizon region (extending over radii of a few $M$) of black holes with infinite throats.
When applied to a (near-)extreme Kerr black hole, and more precisely its (near-)NHEK geometry, this proposal goes under the name of ``Kerr/CFT conjecture'' \cite{Guica:2008mu}.
We emphasize that this correspondence is much more conjectural than the ones discussed so far, as it has no microscopic foundation to rest upon.
Here, Kerr is the analogue of the black brane supergravity solution (with NHEK the analogue of its AdS throat), but we do not know how to build it from weakly interacting branes that backreact at strong coupling to collapse into a black hole.
If we did, then Kerr/CFT would be ``derivable'' from the top-down approach applied to this microscopic picture, and enjoy the same status as AdS/CFT.

In the absence of such a top-down approach, one can work bottom-up using symmetries and scattering experiments.
As we reviewed in Sec.~\ref{subsec:ExtremeKerr}, in the near-extremal regime, a Kerr black hole also develops a near-horizon AdS-like throat geometry, with an emergent conformal symmetry realized as an $\mathsf{SL}(2,\mathbb{R})$ isometry group of the throat.
Moreover, an asymptotic symmetry group analysis reveals that this global $\mathsf{SL}(2,\mathbb{R})$ symmetry is extended to a copy of the full Virasoro algebra \cite{Guica:2008mu}.
This observation is a prediction from GR and forms the basis for the Kerr/CFT correspondence, which conjectures that physics in the near-extreme Kerr throat is dual to a CFT$_2$ living on the boundary (again, this is the picture in the right panel of Fig.~\ref{fig:Throat}, with the CFT on the green cut).
Evidence for this conjecture can also be obtained by scattering waves off the black hole, and noticing (once again) that contributions from the near-horizon throat geometry also take the form of CFT$_2$-like correlators, albeit ones with complex conformal weights \cite{Bredberg:2009pv}.
This last part precludes the theory from being unitary (and hence a bona fide CFT), but then again, the NHEK region connects to the asymptotically flat far region, and is therefore an open (non-unitary) system.

We emphasize that the symmetry and scattering bases for the Kerr/CFT correspondence are \textit{not} conjectured: both aspects are derivable purely within GR (in line with the bottom-up approach).
What is conjectural is that there exists an underlying exact top-down duality: that is an assumption.

\subsection{The photon ring in holography}

Much has yet to be understood about the Kerr/CFT correspondence, including the precise nature of the dual holographic CFT.
It would also be interesting to extend the conjecture to non-extreme Kerr black holes, though as discussed before, the absence of a throat region with a boundary and decoupling limit makes this much more difficult away from extremality.

A speculative idea put forth in \cite{Hadar:2022xag} is that the Kerr photon shell plays a role in the holographic description of a generic non-extreme Kerr black hole, and may be part of its hologram.
There is no single strong argument to suggest this; only small, suggestive observations.
Adopting a bottom-up perspective, for instance, \cite{Hadar:2022xag} found (in the symmetry approach) that the Kerr photon shell has an emergent conformal symmetry generalizing that of NHEK, and (in the scattering approach) that this $\mathsf{SL}(2,\mathbb{R})$ acts upon the ringdown spectrum of Kerr QNMs, which are poles in the two-point function of the black hole, and hence of its holographic dual (Ruelle resonances).

More precisely, the Kerr photon shell is the unique region of geodesic phase space that is invariant under the action of an $\mathsf{SL}(2,\mathbb{R})$ group whose dilations push geodesics closer to this critical surface, physically making the rays in spacetime spend more time orbiting around the black hole \cite{Hadar:2022xag}.
For a near-extreme black hole, this $\mathsf{SL}(2,\mathbb{R})$ action is geometrically realized as an isometry group of the near-horizon AdS-like throat (the basis for Kerr/CFT), whose image forms part of the Kerr critical curve: the NHEKline.
But the NHEKline is only one part of the critical curve, and one may think that the rest also forms an image of the holographic plate---after all, as the above quote from \cite{Maldacena:1996ix} points out, it is possible for a CFT to be dual to more than just the near-horizon throat\footnote{Asking how much of the region outside the horizon is included in the hologram is akin to asking, in the analysis of the scattering problem via matched asymptotic expansions, at which precise radius $r_c$ the matching between the near and far solutions is made.
This question does not have a sharp answer: the matching occurs over an entire region of overlap, which depends on the energy of the modes being matched. That is, the region of overlap is not only a region of spacetime but in fact a region of phase space (like the photon shell).}.

Of course, these are pure speculations.
In order to substantiate these ideas, we would like to test aspects of this picture.
A simple question is whether the photon ring could be part of the hologram dual to Kerr?

To answer this question, we would like to 1) show that the photon ring is encoded in the (eikonal part of the) QNM spectrum, and 2) show that the QNM spectrum is encoded in the holographic theory as its Ruelle resonances of the dual CFT.
In Kerr, \cite{Hadar:2022xag} did 1), but part 2) seems intractable.
In the context of a black hole embedded in AdS, part 2) is tractable in various regimes, but the issue there is that part 1) does not hold, essentially because the presence of the AdS boundary affects the QNM boundary conditions---physically, there can be long-lived bound states formed between the peak of the geodesic potential and the boundary, and these affect the late-time ringdown \cite{Dodelson:2020lal,Dodelson:2022eiz,Dodelson:2023nnr}.
At the same time, the answer according to AdS/CFT must be that the entire spacetime, not just the throat or its near-vicinity, is described by a dual theory living at the AdS boundary.

Nevertheless, in this context, it is interesting to note that the photon ring---which is a universal signature of a black hole---also has a universal signature in the dual CFT, first identified in \cite{Dodelson:2020lal} and further explored in \cite{Dodelson:2023nnr}: the two-point function dual to a black hole in AdS exhibits additional ``bulk-cone singularities'' that are quite mysterious from the boundary perspective, but which can be explained as responses carried from boundary point to boundary point along bulk geodesics orbiting around the black hole a different number of times.
Such paths add new singularities to the usual one along the past and future light cones.
From the boundary perspective, these singularities may appear surprising, but the bulk picture explains their origin and guarantees that they cannot carry information superluminally.

A slightly simpler problem to tackle is to address this last question in the context of a toy model of lower-dimensional gravity.
In other words, is there some toy model of a black hole which does have a photon ring and for which we could prove that it is encoded in the hologram?

The first candidate to come to mind is the three-dimensional BTZ black hole, for which \cite{Birmingham:2001hc, Birmingham:2001pj} already computed the exact QNM spectrum and matched it the Ruelle resonances of a CFT$_2$ (part 2 above).
Unfortunately, this does not settle the question because 1) cannot be done, since there is no photon ring.

Another candidate is the warped version of the BTZ black hole, which does have a photon ring, and for which there is hope to independently compute the spectrum of Ruelle resonances in the holographic dual.
Here, we carry out step 1) by showing that in warped AdS$_3$, as in Kerr, the photon ring is a geometrization of the eikonal QNM spectrum.
In this paper, we do not carry out part 2) for the warped AdS$_3$, but we suspect it may be doable via the future investigations of Warped CFTs, along the lines of \cite{Song:2017czq}.
Another advantage is that warped AdS$_3$ is extremely similar to NHEK.
Thus, we regard it as a stepping stone towards tackling the full Kerr black hole directly.

\clearpage

\section{Warped \texorpdfstring{AdS$_3$}{AdS3} black holes}
\label{sec:WBTZ}

Black hole quasinormal modes (QNMs) are solutions of the massless wave equation that are excited when the black hole ``rings down'' in response to external perturbations.
They are obtained by imposing a particular set of boundary conditions that pick out ``resonances'': QNMs must be purely ingoing at the horizon and outgoing at infinity.
In this paper, we compute the scalar QNMs in a Warped AdS$_3$ (WAdS$_3$) black hole background and show that, in the high-frequency (or ``eikonal'') limit, they are encoded in the physics of the black hole photon ring.

As reviewed in Sec.~\ref{subsec:Geometrization}, the spectrum of eikonal QNMs of a Kerr black hole may be expressed in terms of the angular velocity and Lyapunov exponent of the Kerr bound photon orbits [Eq.~\eqref{eq:KerrQNMs}].
This relation was recently extended to the case of self-dual warped AdS$_3$ \cite{Kapec:2022dvc}.
However, it does not generally hold for black holes in non-asymptotically-flat spacetimes, where the imposition of QNM boundary conditions requires more care.
For instance, for asymptotically AdS black holes, the simple relation \eqref{eq:KerrQNMs} breaks down because the presence of the AdS boundary affects the QNM spectrum, and there exist long-lived resonances between the black hole and boundary \cite{Festuccia:2008zx,Berti:2009kk}.

For asymptotically flat or AdS spacetimes, there is a well-defined notion of asymptotic boundary, which may be defined via the process of conformal completion.
In the case of a WAdS$_3$ black hole, the nature of the asymptotic boundary is more subtle, in particular because the spacetime appears not to be conformally complete \cite{Skenderis:2009nt,Horava:2010vho}. As a result, the proper definition of QNM boundary conditions is not clear.
In this section, we will identify a choice of ``resonant'' boundary conditions that provide a reasonable definition of QNMs, such that the familiar relation \eqref{eq:KerrQNMs} from Kerr between the eikonal QNM spectrum and the black hole photon ring continues to hold.

The WAdS$_3$ black hole under consideration appeared in \cite{Nutku:1993eb,Gurses:1994bjn,Moussa:2003fc,Bouchareb:2007yx,Compere:2007in} and was first studied in the context of holography in \cite{Anninos:2008fx}.
In units where the AdS radius is $l=1$, the metric is given by
\begin{subequations}
\label{eq:WBTZ}
\begin{align}
    ds^2&=-N(r)^2\ed t^2+\frac{\ed r^2}{4R(r)^2N(r)^2}+R(r)^2\left(\ed\theta+N^{\theta}(r)\ed t\right)^2\,,\\
	R(r)^2&=\frac{r}{4}\left(3\left(\nu^2-1\right)r+\left(\nu^2+3\right)(r_++r_-)-4\nu\sqrt{r_+r_-\left(\nu^2+3\right)}\right)\,,\\
	N(r)^2&=\frac{1}{4R(r)^2}(\nu^2+3)(r-r_+)(r-r_-)\,,\\
	N^{\theta}(r)&=\frac{2\nu r-\sqrt{r_+r_-(\nu^2+3)}}{2R(r)^2}\,,
\end{align}
\end{subequations}
where $\theta\in[0,2\pi[$ is an angle and $r\in [0,+\infty[$ is a radial coordinate, while $\nu$ and $r_\pm$ are constant parameters.
The black hole has two horizons (an inner and an outer one) located at $r_+>r_->0$.
One must assume that $\nu^2>1$ in order to avoid closed timelike curves, and we will further restrict our attention to $\nu>1$ for simplicity.
These solutions may be obtained as quotients of spacelike warped AdS$_3$ by a discrete subgroup of the isometry group \cite{Anninos:2008fx}, in much the same way that BTZ black holes are quotients of global AdS$_3$.
In fact, when $\nu\to1$, the metric \eqref{eq:WBTZ} becomes locally AdS$_3$ and reduces to the usual BTZ black hole, albeit in an unusual coordinate system.

In this paper, we will consider the physical behavior of fields at the horizon and at an asymptotic boundary, which we will take to be at $r\to\infty$.
This is justified because this locus corresponds to a boundary in the projection diagram of this black hole \cite{Bieliavsky:2024hus}.
It is also where the charges and asymptotic symmetries of the black hole are defined (on constant-time slices) \cite{Compere:2007in,Compere:2009zj,Blagojevic:2009ek,Henneaux:2011hv}, and the region of anisotropic conformal infinity \cite{Horava:2010vho}.

\subsection{Warped photon ring}
\label{subsec:PhotonRing}

The WAdS$_3$ black hole \eqref{eq:WBTZ} has a photon ring of gravitationally bound light orbits.
That is, there exist affinely parameterized null geodesics $x^\mu(s)=(t(s),r(s),\theta(s))$, with momentum $\dot{x}^\mu=\partial_s x^\mu$, whose radial motion remains bounded outside the horizon.
We now show this explicitly.

\begin{figure}[t!]
    \centering
    \includegraphics[scale=0.6]{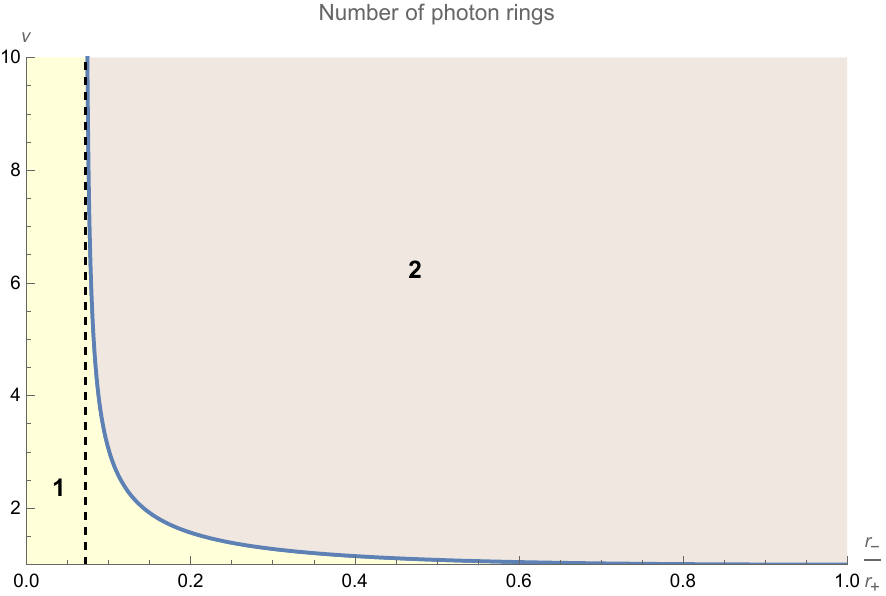}
    \caption{Number of photon rings for the WAdS$_3$ black hole \eqref{eq:WBTZ} as a function of $\nu\in[1,+\infty)$ and the ratio $r_-/r_+\in(0,1]$.
    In the grey region, there are two photon rings $\tilde{r}_\pm$, whereas in the yellow region, there is only one outer photon ring $\tilde{r}_+$ outside the two horizons.
    Along the blue line delineating the boundary between these two regimes, the inner photon ring is located at $\tilde{r}_-=0$.
    This boundary asymptotes to the dashed vertical line at $\frac{r_-}{r_+} =\frac{2-\sqrt{3}}{2+\sqrt{3}}$.}
    \label{fig:PhotonRingNumber}
\end{figure}

The metric \eqref{eq:WBTZ} admits two Killing vector fields $\partial_t$ and $\partial_\theta$ that define two conserved quantities along null geodesics: the energy $E\equiv\partial_t^\mu\dot{x}_\mu=-\dot{x}_t$ and the angular momentum $L\equiv\partial_\theta^\mu\dot{x}_\mu=\dot{x}_\theta$,
\begin{align}
    E&=-\dot{t}+\frac{1}{2}\left(-2r\nu+\sqrt{r_+r_-\left(\nu^2+3\right)}\right)\dot{\theta}\,,\\
    L&=\frac{r}{4}\left[3r\left(\nu^2-1\right)\dot{\theta}-\frac{2}{r}\sqrt{r_+r_-\left(\nu^2+3\right)}\dot{t}+\eta\left(\nu^2+3\right)\dot{\theta}+4\nu\left(\dot{t}-\dot{\theta}\sqrt{r_+r_-\left(\nu^2+3\right)}\right)\right]\,,
\end{align}
where $\eta=r_++r_-$.
Null geodesics are also subject to the condition $g_{\mu\nu}\dot{x}^\mu\dot{x}^\nu=0$ and thus obey
\begin{align}
    \label{eq:RadialMotion}
    \dot{r}^2+V(r)=0\,, 
\end{align}
the radial geodesic equation with potential
\begin{align}
    \label{eq:RadialPotential}
    V(r)=-4\left[L^2+2LER(r)^2N^\theta(r)+E^2R(r)^2\right]\,.
\end{align}
Simultaneously imposing the conditions 
\begin{align}
    V(r)=V'(r)=0\,,
\end{align}
defines null geodesics with fixed radial position $r(s)=\tilde r$.
Such geodesics correspond to photons that orbit around the black hole at fixed radius, forming a photon ring.
Assuming that $E\neq 0$, these two conditions admit two solutions that correspond to photon orbits bound at two radii\footnote{In the special case $E=0$, these two conditions are also fulfilled by $L=0$ at any radius.}
\begin{align}
    \label{eq:OrbitalRadii}
    \tilde{r}_\pm=\frac{\eta}{2}\pm\frac{(r_+-r_-)\nu}{\sqrt{3\left(\nu^2-1\right)}}\,,\qquad
    \eta=r_++r_-\,,
\end{align}
with critical ``specific angular momentum'' $\lambda=\frac{L}{E}$ given by
\begin{align}
    \label{eq:CriticalLambda}
    \tilde{\lambda}_\pm=\frac{1}{2}\left[\sqrt{r_+r_-\left(\nu^2 +3\right)}-\eta\nu\mp\frac{1}{2}(r_+-r_-)\sqrt{3\left(\nu^2-1\right)}\right]
    <0\,.
\end{align}
This critical angular momentum is negative for both orbital radii---this will prove important when we examine the direction of propagation of QNMs near the rings.
The radii $\tilde{r}_\pm$ are ordered as
\begin{align}
    \tilde{r}_+>r_+>r_->\tilde{r}_-\,.
\end{align}
Thus, there may be an outer photon ring at $\tilde{r}_+$ outside the horizon, and an inner one at $\tilde{r}_-$ inside the horizon.
Since the radial coordinate $r$ cannot become negative \cite{Anninos:2008fx}, both are present only when
\begin{align}
    \frac{r_-}{r_+}\geq\frac{2\nu-\sqrt{3\left(\nu^2-1\right)}}{2\nu+\sqrt{3\left(\nu^2-1\right)}}\,.
\end{align}
This condition remains nontrivial even in the limit of large $\nu\to\infty$, where the lower bound tends to a nonzero value.
At the other extreme, in the unwarped limit $\nu\to1$ where the black hole \eqref{eq:WBTZ} reduces to the usual BTZ solution, the only way to have two photon rings is to take the extremal limit in which the two horizons coincide.
Provided that
\begin{align}
    \label{eq:nuCondition}
    \frac{r_-}{r_+}\ge\frac{2-\sqrt{3}}{2+\sqrt{3}},
\end{align}
this condition may also be expressed in terms of $\nu$ as 
\begin{align}
    \nu\geq\nu_{\rm c}\equiv\sqrt{\frac{3\eta^2}{14r_+r_--r_+^2-r_-^2}}
    \ge1\,.
\end{align}
Since we allow $\nu\ge1$, this implies that there is a critical value $\nu=\nu_{\rm c}$ for which the inner photon ring disappears.
On the other hand, when Eq.~\eqref{eq:nuCondition} is not satisfied, there is always a single photon ring for any value of $\nu$, as illustrated in Fig.~\ref{fig:PhotonRingNumber}.
When $\nu=\nu_{\rm c}$, the inner photon ring is located at $\tilde r_-=0$ and its critical parameter vanishes:
\begin{align}
    \label{eq:ZeroMomentum}
    \nu=\sqrt{\frac{3\eta^2}{14r_+r_--r_+^2-r_-^2}}
    \qquad\Longrightarrow\qquad
    \tilde{\lambda}_-=0\,.
\end{align}
The photon orbits at radii $\tilde{r}_\pm$ have angular velocities $\tilde{\Omega}_\pm$ and half-orbital periods $\tau_\pm$ given by \footnote{Here, an orbit is one full rotation around the black hole, but we instead count the half-orbit number $n$, following the convention in Kerr where $n$ labels successive images of the photon ring \cite{Johnson:2019ljv}.}
\begin{align}
    \label{eq:CriticalOmega}
    \tilde{\Omega}_\pm=\left.\frac{d\theta(s)}{dt(s)}\right|_{\tilde{r}_\pm,\tilde{\lambda}_\pm}
    =\frac{1}{\tilde{\lambda}_\pm}
    <0\,,\qquad
    \tau_\pm=-\frac{\pi}{\tilde{\Omega}_\pm}
    =-\pi\tilde{\lambda}_\pm
    >0\,.
\end{align}
The photon rings are unstable because at the critical orbital radii, 
\begin{align}
    V''(\tilde{r}_\pm)=-6E^2\left(\nu^2-1\right)
    <0\,.
\end{align}

We now define deviations from the photon ring orbital radius and critical angular momentum
\begin{align}
    \label{eq:NearRingVariation}
    \delta r=r-\tilde{r}_\pm\,,\qquad
    \delta\lambda=\lambda-\tilde{\lambda}_\pm\,.
\end{align}
A precisely bound photon has $\delta r=0$ (a spacetime condition) and $\delta\lambda=0$ (a momentum-space condition).
Together, these conditions define the photon rings as the phase-space locus
\begin{align}
    \text{PHOTON RING:}\quad
    \delta r=\delta\lambda=0\,.
\end{align}
Consider now two nearby geodesics, one that is exactly bound at $\tilde{r}_\pm$, and another that is initially offset by an infinitesimal radial separation $|\delta r_0|\ll1$.
By Eq.~\eqref{eq:RadialMotion}, its geodesic deviation grows as
\begin{align}
    \delta r(s)\approx\delta r_0e^{\sqrt{-\frac{1}{2}V''(\tilde{r}_\pm)}s}\,,
\end{align}
for as long as it remains in the vicinity of the ring, or more precisely, so long as it remains in the
\begin{align}
    \label{eq:NearRingRegion}
    \text{NEAR-RING REGION:}\quad
    \begin{cases} 
        |\delta r|\ll\eta
        &\text{(near-peak)}\,,\\
        \left|\delta\lambda\right|\ll\eta
        &\text{(near-critical)}\,.
    \end{cases}
\end{align}
The first condition zooms into the spacetime of the bound photon orbit while the second condition zooms into the bound orbit in momentum space.
After $n=\frac{\Delta\theta}{\pi}$ half-orbits, the separation grows to
\begin{align}
    \delta r(n)\approx\delta r_0e^{\gamma n}\,,
\end{align}
where $\gamma$ is the Lyapunov exponent controlling the orbital instability of the photon ring,
\begin{align}
    \gamma\equiv\sqrt{-\frac{1}{2}V''(\tilde{r}_\pm)}\left|\frac{\pi}{\dot{\theta}}\right|_{\tilde{r}_\pm,\tilde{\lambda}_\pm}\,.
\end{align}
In the warped AdS$_3$ black hole spacetime, the two photon rings share the same Lyapunov exponent
\begin{align}
    \gamma=\frac{\pi}{4}(r_+-r_-)\left(\nu^2+3\right)\,.
\end{align}
The period-averaged radial deviation as a function of time $t$ can thus be expressed as
\begin{align}
    \delta r(t)=\delta r_0e^{\gamma_Lt}\,,
\end{align}
where $\gamma_L$ is the ``Lyapunov exponent in time'' (rather than with respect to orbits)
\begin{align}
    \gamma_L\equiv\frac{\gamma}{\tau}
    =\sqrt{-\frac{1}{2}V''(\tilde{r}_\pm)}\left|\frac{1}{\dot t}\right|_{\tilde{r}_\pm,\tilde{\lambda}_\pm}\,.
\end{align}
Intuitively, this relation makes sense because $\gamma_L\sim\frac{dr}{dt}=\frac{dr}{dn}\cdot\frac{dn}{dt}\sim\gamma\cdot\frac{1}{\tau}$.
Unlike the orbital Lyapunov exponent, $\gamma_L$ differs for the two photon rings:
\begin{align}
    \label{eq:LyapunovInTime}
    \gamma_{L\pm}=-\frac{\nu^2+3}{4}(r_+-r_-)\tilde{\Omega}_\pm\,
    >0.
\end{align}
Together with $\theta\approx\tilde{\Omega}_\pm t$, these equations provide a complete description of null geodesic motion for nearly bound photons in the near-ring region of the warped AdS$_3$ black hole.

\subsection{Exact spectrum of warped QNMs}
\label{subsec:WarpedQNMs}

In the geometric-optics limit, wavefronts are approximated by null geodesic congruences, and the effects of spin are subleading.
As such, even though we are ultimately interested in photons or gravitons propagating around the WAdS$_3$ black hole \eqref{eq:WBTZ}, it is sufficient for our purposes to consider only a massless scalar in that background.
Scalar QNMs are solutions of the massless wave equation
\begin{align}
    \label{eq:Laplacian}
    \nabla^2\Phi(t,\theta,r)=0\,.
\end{align}
that are also eigenfunctions of the operators $\partial_t$ and $\partial_\theta$ (the Killing vector fields generating the translational isometries along $t$ and $\theta$) and which obey specific boundary conditions (ingoing at the horizon and outgoing at infinity).
These eigenmodes take the form
\begin{align}
    \label{eq:ModeAnsatz}
    \Phi(t,\theta,r)=e^{-i\omega t+ik\theta}\phi(r)\,,
\end{align}
where $\omega$ denotes the energy of the scalar mode and $k$ its (quantized) angular momentum.
Under this ansatz, the scalar wave equation \eqref{eq:Laplacian} reduces to a radial Schr\"odinger-type equation
\begin{align}
    \label{eq:RadialODE}
    \left[\rho\partial_r(\rho\partial_r)+V_{\rm QNM}(r)\right]\phi(r)=0\,,
\end{align}
where we introduced a new radial function
\begin{align}
    \rho(r)=\left(\nu^2+3\right)(r-r_+)(r-r_-)
    =4R(r)^2N(r)^2\,,
\end{align}
together with a radial wave potential
\begin{align}
    \label{eq:WavePotential}
    V_{\rm QNM}(r)=4\left[k^2+2k\omega R(r)^2N^\theta(r)+\omega^2R(r)^2\right]\,.
\end{align}
It is also useful to introduce a new radial coordinate
\begin{align}
    z=\frac{r-r_+}{r-r_-}\,, 
\end{align}
in terms of which the horizon and spatial infinity lie at $z=0$ and $z=1$, respectively, and the radial ODE \eqref{eq:RadialODE} takes the simpler form
\begin{align}
    z(1-z)\phi''(z)+(1-z)\phi'(z)-\left[\frac{A^2}{z}-B^2+\frac{C(C-1)}{1-z}\right]\phi(z)=0\,,
\end{align}
with
\begin{subequations}
    \begin{align} 
        A&=\frac{i}{(r_+-r_-)\left(\nu^2+3\right)}\left[2k+\omega\left(2\nu r_+-\sqrt{r_+r_-\left(\nu^2+3\right)}\right)\right]\,,\\ 
        B&=\frac{i}{(r_+-r_-)\left(\nu^2+3\right)}\left[2k+\omega\left(2\nu r_--\sqrt{r_+r_-\left(\nu^2+3\right)}\right)\right]\,,\\
        C&=\frac{1}{2}\left(1-\sqrt{1-\frac{12\left(\nu^2-1\right)\omega^2}{\left(\nu^2+3\right)^2}}\right)\,.
\end{align}
\end{subequations}
This linear second order differential equation can be brought into the standard form of the hypergeometric differential equation via the function redefinition \cite{Chen:2009rf} $\phi(z)=z^A(1-z)^Cf(z)$, such that $f(z)$ satisfies the standard hypergeometric differential equation \cite{Whittaker_Watson_1996}.

The general mode solution is thus a sum of two modes
\begin{align}
    \label{eq:GeneralMode}
    \Phi(t,\theta,z)&=C_+\Phi_+(t,\theta,z)+C_-\Phi_-(t,\theta,z)\,,\\
    \Phi_\pm(t,\theta,z)&=e^{-i\omega t+ik\theta}(1-z)^Cz^{\pm A}{_2F_1}(\pm A+B+C,\pm A-B+C;1\pm 2A;z)\,,
\end{align}
where ${_2F_1}(a,b;c;z)$ denotes Gauss' hypergeometric function \cite{olver10}.

With the general solution to the massless wave equation in hand, our next objective is to impose suitable boundary conditions to make them behave like the Kerr QNMs in the eikonal limit.
This requires us to first analyze the near-horizon behavior of these two modes.
Since the hypergeometric function approaches unity as $z\to0$, near the horizon we may approximate them as
\begin{align}
    \Phi_\pm(t,\theta,z)\stackrel{z\to0}{\approx}e^{-i\omega t+ik\theta}(1-z)^Cz^{\pm A}\left[1+\text{(subleading terms)}\right]\,.
\end{align}
To proceed, we first rewrite
\begin{align}
    \label{eq:A}
    A=ic_1k+ic_2\omega\,,
\end{align} 
where
\begin{align}
    c_1=\frac{2}{(r_+-r_-)(\nu^2+3)}>0\,,\qquad
    c_2=\frac{2\nu r_+-\sqrt{r_+r_-(\nu^2+3)}}{(r_+-r_-)(\nu^2+3)}>0\,.
\end{align}
Next, since QNM frequencies are complex, we decompose them into real and imaginary parts
\begin{align}
    \omega=\omega_R-i\omega_I\,.
\end{align}
Then, we may rewrite the near-horizon behavior of the modes $\Phi_\pm(t,\theta,z)$ as
\begin{align}
    \Phi_\pm(t,\theta,z)\stackrel{z\to0}{\approx}e^{-i\omega_R\left[t\mp\left(c_1\frac{k}{\omega_R}+c_2\right)\ln{z}\right]+ik\theta}e^{-\omega_It+c_2\omega_I\ln{z}}\,,
\end{align}
where the first exponential is a pure phase (a traveling wave) while the second exponential is a dissipative term (a damped exponential for $\omega_I>0$).

The direction of propagation of these two modes near the horizon is determined by the sign of $c_1\frac{k}{\omega_R}+c_2$.
We will be interested in QNMs whose frequencies in the eikonal limit of high frequency and large momentum behave like their Kerr analogues \eqref{eq:KerrQNMs}, that is, for which
\begin{align}
    \omega_R\approx{\tilde{\Omega}_\pm}k
    =\frac{k}{\tilde{\lambda}_\pm}\,.
\end{align}
For such modes,
\begin{align}
    c_1\frac{k}{\omega_R}+c_2\approx c_1{\tilde\lambda}_\pm+c_2>0\,,
\end{align}
which implies that $\Phi_+$ is an outgoing mode at the horizon, whereas $\Phi_-$ is ingoing.
Therefore, to obtain purely ingoing (or purely outgoing) solutions, we must impose $C_+=0$ (respectively, $C_-=0$) in the general solution \eqref{eq:GeneralMode}.

We now examine the horizon flux of these modes.
The flux of a complex scalar field is \cite{Chen:2009rf} 
\begin{align}
    \mathcal F=\frac{\sqrt{-g}\,g^{rr}}{2i}\left(\Phi^*\partial_r\Phi-\Phi\partial_r\Phi^*\right)
    =\frac{\nu^2+3}{4i}(r_+-r_-)z\left(\Phi^*\partial_z\Phi-\Phi\partial_z\Phi^*\right)\,.
\end{align}
Near the horizon, the flux of the general solution \eqref{eq:GeneralMode} becomes proportional to
\begin{align}
    \mathcal F\stackrel{z\to0}{\propto}A_R\left(|C_+|^2z^{2c_2\omega_I}+|C_-|^2z^{-2c_2\omega_I}\right)-ic_2\omega_I\left(C_-^*C_+z^{2iA_R}-C_+^*C_-z^{-2iA_R}\right)\,,
\end{align}
where $A_R$ denotes the real part of $A$ in Eq.~\eqref{eq:A}.
We observe from this expression that a purely ingoing mode with $C_+ = 0$ has a divergent flux at the horizon if $\omega_I$ is positive, as will be the case in the eikonal limit [Eq.~\eqref{eq:EikonalCase1} below].
On the other hand, a purely outgoing mode with $C_- = 0$ retains a finite flux through the horizon.

Next, we analyze the asymptotic behavior of the general solution \eqref{eq:GeneralMode} near spatial infinity.
Using a standard formula for the hypergeometric function, we may expand it around $z=1$ as
\begin{align}
    \Phi(t,\theta,z)\approx e^{-i\omega t+ik\theta}\left[Q_+(1-z)^{\frac{1}{2}+\varpi}+Q_-(1-z)^{\frac{1}{2}-\varpi}\right]\,,
\end{align}
where for clarity, we introduced
\begin{align}
    \label{eq:b}
    \varpi\equiv\frac{1}{2}-C
    =\sqrt{\frac{1}{4}-\frac{3\left(\nu^2-1\right)\omega^2}{(\nu^2+3)^2}}
    =\sqrt{\frac{1}{4}-\tilde{b}\omega^2}\,,\qquad
    \tilde{b}=\frac{3\left(\nu^2-1\right)}{(\nu^2+3)^2}\,,
\end{align}
and the coefficients $Q_\pm$ and $C_\pm$ are related by the connection formulas 
\begin{align}
    Q_+&=C_+\frac{\Gamma(1+2A)\Gamma(2C-1)}{\Gamma(A+B+C)\Gamma(A-B+C)}+C_-\frac{\Gamma(1-2A)\Gamma(2C-1)}{\Gamma(-A+B+C)\Gamma(-A-B+C)}\,,\\
    Q_-&=C_+\frac{\Gamma(1+2A)\Gamma(1-2C)}{\Gamma(1+A+B-C)\Gamma(1+A-B-C)}+C_-\frac{\Gamma(1-2A)\Gamma(1-2C)}{\Gamma(1-A+B-C)\Gamma(1-A-B-C)}\,.
\end{align}
Since $\varpi$ is complex, we may further decompose it as $\varpi=\varpi_R+i\varpi_I$.
In the eikonal limit, we have
\begin{align}
    \varpi\approx\varpi_R\approx i\sqrt{\tilde{b}}\omega_R\,,
\end{align}
where for a complex number $z$, we define $\sqrt{z}$ to be the square root with positive real part (or for purely imaginary $z$, the square root with positive imaginary part).

Therefore, at spatial infinity in the eikonal limit, the general solution \eqref{eq:GeneralMode} behaves as
\begin{align}
    \Phi(t,\theta,z)\approx(1-z)^\frac{1}{2}e^{ik\theta}\left(Q_+e^{-i\omega_R\left[t-\sqrt{\tilde{b}}\ln(1-z)\right]}+Q_-e^{-i\omega_R\left[t+\sqrt{\tilde{b}}\ln(1-z)\right]}\right)\,.
\end{align}
Since the function $\ln(1-z)$ is decreasing as $z\to1$, we conclude that the mode proportional to $Q_+$ is ingoing at infinity while the mode proportional to $Q_-$ is outgoing at infinity.

Near infinity, the flux becomes proportional to
\begin{align}
    \mathcal{F}&\stackrel{z\to1}{\propto}i\varpi_I\left[|Q_-|^2(1-z)^{-2\varpi_R}-|Q_+|^2(1-z)^{2\varpi_R}\right]\nonumber\\
    &\phantom{\stackrel{z\to1}{\propto}}-\varpi_R\left[Q_-^*Q_+(1-z)^{2i\varpi_I}-Q_+^*Q_-(1-z)^{-2i\varpi_I}\right]\,.
\end{align}
Since $\varpi_R>0$, an outgoing wave (with $Q_+=0$) has divergent flux, whereas an ingoing wave (with $Q_-=0$) has finite flux.
We could also impose a Dirichlet boundary condition at infinity
\begin{align}
     \Phi(t,\theta,z)\stackrel{z\to 1}{\longrightarrow}0\,.
\end{align}
This would not impose a condition on $Q_+$, since the real part of $\frac{1}{2} + \varpi$ is always positive.
As such, the Dirichlet condition only implies that $Q_-=0$, just like the finite-flux condition.

In summary, the usual QNM boundary conditions, corresponding to ingoing waves at the horizon and outgoing waves at infinity, would require that we set $C_+=0$ and $Q_+=0$.
In an asymptotically flat spacetime, these boundary conditions lead to finite fluxes at both horizon and infinity, and unambiguously define the QNMs.
On the other hand, in the WAdS$_3$ black hole spacetime, these boundary conditions lead to a divergent flux at infinity, and do not match the boundary conditions of \cite{Chen:2009rf, Martin:2022duk}, which instead demanded ingoing waves at the horizon ($C_+=0$) and a finite flux at infinity ($Q_-=0$).
In this setting, both choices for how to define QNMs appear equally reasonable, so we will now examine each possibility (outgoing waves or finite flux at infinity) in turn---in the end, we will see that these two types of ``QNMs'' are associated with the two different photon rings.

\subsubsection{Outgoing boundary conditions at infinity}

For both choices of boundary conditions, we require the waves to be purely ingoing at the horizon.
To also make them purely outgoing at infinity (the usual QNM definition), we must set $Q_+=0$.
This can be achieved in one of two ways: letting $n$ be a positive integer, we may demand either
\begin{align}
    -A+B+C=-n\qquad\text{or}\qquad
    -A-B+C=-n\,.
\end{align}

\paragraph{Case 1: $-A-B+C=-n$.} 
This condition leads to 
\begin{align}
    \label{eq:Case1}
    \frac{1}{2}\left(1-\sqrt{1-\frac{12\left(\nu^2-1\right)\omega^2}{\left(\nu^2+3\right)^2}}\right)-\frac{i(4k+\omega\delta)d}{\left(\nu^2+3\right)}=-n\,,
\end{align}
where we have introduced 
\begin{align}
    \delta=2\left[\nu(r_++r_-)-\sqrt{r_+r_-(\nu^2+3)}\right],\qquad
    d=\frac{1}{r_+-r_-}\,.
\end{align}
To solve this condition for the frequency $\omega$, one must isolate the square root and then square both sides of the equation, which introduces a spurious root; one must then take care to single out the true root at the end.
The correct solution to Eq.~\eqref{eq:Case1} is given by
\begin{align}
    \label{eq:QNMsCase1}
    \omega_{1+}=-\frac{\delta}{4}\Xi+\frac{\sign(k)}{4}\sqrt{3\left(\nu^2-1\right)(r_+-r_-)^2\Xi^2-4\gamma_{L+}\gamma_{L-}}\,,
\end{align}
where
\begin{align}
    \Xi=\tilde{\Omega}_+\tilde{\Omega}_-\left[k+\frac{i}{4}\left(n+\frac{1}{2}\right)\left(\nu^2+3\right)(r_+-r_-)\right]\,,
\end{align}
which can be rewritten as
\begin{align}
    \Xi=\tilde{\Omega}_+\left[\tilde{\Omega}_-k-i\left(n+\frac{1}{2}\right)\gamma_{L-}\right]
    =\tilde{\Omega}_-\left[\tilde{\Omega}_+k-i\left(n+\frac{1}{2}\right)\gamma_{L+}\right]\,.
\end{align}
The frequencies \eqref{eq:QNMsCase1} exhibit a symmetry under interchange of the parameters of the two photon rings, but it is really the sign in front of the square root that determines their behavior in the eikonal regime of large $k$, and whether the modes are ultimately associated to the inner or outer photon ring in that limit. 
For these frequencies, the eikonal limit is
\begin{align}
    \label{eq:EikonalCase1}
    \omega_{1+}\stackrel{|k|\gg1}{\approx}\tilde{\Omega}_+k-i\left(n+\frac{1}{2}\right)\gamma_{L+}\,.
\end{align}

\paragraph{Case 2: $-A+B+C=-n$.}
This condition leads to
\begin{align}
    \label{eq:Case2}
     \frac{1}{2}\left(1-\sqrt{1-\frac{12\left(\nu^2-1\right)\omega^2}{\left(\nu^2+3\right)^2}}\right)-\frac{2i\omega\nu}{\left(\nu^2+3\right)}=-n\,.
\end{align}
In this case, the correct solution for $\omega$ is
\begin{align}
    \label{eq:QNMsCase2}
    \omega_{2+}=-i\nu(2n+1)+i\sqrt{3\left(\nu^2-1\right)n(n+1)+\nu^2}\,.
\end{align}
These frequencies are purely imaginary and are independent of the angular momentum $k$, but this statement depends on the choice of coordinates---compare with Eq.~\eqref{eq:NotImaginary} in Appendix~\ref{app:QuadraticEnsemble}.

\subsubsection{Finite-flux and Dirichlet boundary conditions at infinity}

Another condition that one could naturally impose at infinity to define ``QNMs'' (as was done in \cite{Chen:2009rf, Martin:2022duk}) is that the flux remain finite there.
As mentioned before, this constraint is equivalent to a Dirichlet boundary condition at infinity.
Interestingly, the resulting frequencies are the spurious ones encountered in the last section when squaring the square roots.

\paragraph{Case 1: $1-A-B-C=-n$.}
This condition leads to 
\begin{align}
    \label{eq:Case3}
    \frac{1}{2}\left(1+\sqrt{1-\frac{12\left(\nu^2-1\right)\omega^2}{\left(\nu^2+3\right)^2}}\right)-\frac{i(4k+\omega\delta)d}{\left(\nu^2+3\right)}=-n\,.
\end{align}
In this case, the correct solution (which would have been the spurious one in the last section) is
\begin{align}
    \omega_{1-}=-\frac{\delta}{4}\Xi-\frac{\sign(k)}{4}\sqrt{3\left(\nu^2-1\right)(r_+-r_-)^2\Xi^2-4\gamma_{L+}\gamma_{L-}}\,,
\end{align}
with eikonal limit
\begin{align}
    \label{eq:EikonalCase3}
    \omega_{1-}\stackrel{|k|\gg1}{\sim}\tilde{\Omega}_-k-i\left(n+\frac{1}{2}\right)\gamma_{L-}\,.
\end{align}
These frequencies depend on the parameters of the inner photon ring instead of the outer one.

\paragraph{Case 2: $1-A+B-C=-n$.}
This condition leads to
\begin{align}
    \frac{1}{2}\left(1+\sqrt{1-\frac{12\left(\nu^2-1\right)\omega^2}{\left(\nu^2+3\right)^2}}\right)-\frac{2i\omega\nu}{\left(\nu^2+3\right)}=-n\,.
\end{align}
This time, the solution is
\begin{align}
    \omega_{2-}=-i\nu(2n+1)-i\sqrt{3\left(\nu^2-1\right)n(n+1)+\nu^2}\,.
\end{align}
Once again, these modes are purely imaginary.

In conclusion, if one requires waves that are purely ingoing at the horizon to also be purely outgoing at infinity (the usual QNM definition), then the resulting modes have the eikonal spectrum
\begin{align}
\label{eq:OuterQNMs}
\begin{split}
    \omega_{1+}&\approx\tilde{\Omega}_+k-i\left(n+\frac{1}{2}\right)\gamma_{L+}\,,\\
    \omega_{2+}&=-i\nu(2n+1)+i\sqrt{3\left(\nu^2-1\right)n(n+1)+\nu^2}\,.    
\end{split}
\end{align}
By contrast, imposing a finite-flux or Dirichlet condition at infinity yields the eikonal spectrum
\begin{align}
\label{eq:InnerQNMs}
\begin{split}
    \omega_{1-}&\approx\tilde{\Omega}_-k-i\left(n+\frac{1}{2}\right)\gamma_{L-}\,,\\
    \omega_{2-}&=-i\nu(2n+1)-i\sqrt{3\left(\nu^2-1\right)n(n+1)+\nu^2}\,.
\end{split}
\end{align}
For each choice of boundary condition at infinity, we found a set of modes with purely imaginary frequencies that depend solely on the parameter $\nu$, and not the angular momentum $k$.
The eikonal limit is therefore not defined for these modes, and we will focus in the following on the other set of resonant modes.

To summarize, we defined two sets of ``resonant'' modes for WAdS$_3$ black holes by imposing two different types of boundary conditions at infinity.
For Kerr black holes, these two different choices coincide and unambiguously define QNMs.
For WAdS$_3$ black holes, either choice seems equally valid, so we considered both of them.
In each case, the resulting ``resonances'' contain modes with no eikonal limit and a set of modes associated with a photon ring: either the outer one (for one choice of boundary condition) or the inner one (for the other choice).
In the following sections, we will use different approaches to recover these eikonal spectra directly: first by taking a geometric (Penrose) limit of the spacetime that scales immediately into the eikonal regime (Sec.~\ref{subsec:PenroseLimit}), then by identifying an underlying conformal symmetry of these spectra (Sec.~\ref{subsec:EikonalSymmetry}), and finally by connecting these resonances to null geodesic congruences in the geometric-optics approximation (Sec.~\ref{subsec:GeometricOptics}).
As expected, all these approaches will turn out to be consistent with each other, recovering the eikonal limit of the exact spectra derived in this section.

\subsection{Penrose limit}
\label{subsec:PenroseLimit}

The Penrose limit \cite{Penrose1976,blau_2011} assigns a plane-wave metric to the region of spacetime around a null geodesic.
In particular, it can be used to describe the geometry around the photon rings \cite{Fransen:2023eqj}.
It was observed in \cite{Fransen:2023eqj} that for self-dual WAdS$_3$, as well as for Schwarzschild and Kerr black holes, it is possible to directly obtain the eikonal QNM spectrum by first taking a Penrose limit that zooms into the photon ring in spacetime, and then solving the wave equation in the resulting limiting geometry (rather than solving the full wave equation first and then taking the eikonal limit via the geometric approximation).
This is summarized by the following diagram (reproduced from \cite{Fransen:2023eqj}):
\begin{figure}[h] 
\begin{center}
\begin{tikzcd}
	\phantom{test}\text{General spacetime} \phantom{t} (M,g) \phantom{test} \arrow[r, "\text{Penrose limit}{}_{\gamma}"] \arrow[d," "]& \phantom{test}\text{Plane wave spacetime} \phantom{t} (M_{\gamma},g_{\gamma})\phantom{test} \arrow[d," "] \\
	\phantom{test}\text{Wave equation on $(M,g)$}\phantom{test} \arrow[r, "\text{Geometric optics}{}_{\gamma}"'] & \phantom{test}\text{Wave equation on $(M_{\gamma},g_{\gamma})$}\phantom{test}
\end{tikzcd}
\end{center}
\caption{In a spacetime $M$ with metric $g$, the leading-order expansion of the full wave equation in the geometric-optics approximation based on the null geodesic $\gamma$ is the exact wave equation on the Penrose-limit spacetime $M_{\gamma}$ with metric $g_{\gamma}$.}
\end{figure}

\subsubsection{Penrose limit in Fermi coordinates}

In this section, we use Fermi coordinates to derive the Penrose-limit geometry around the photon rings of WAdS$_3$ black holes, and show that it takes the form
\begin{align}
    \label{eq:WarpedPenrose}
	ds^2=2\ed u\ed v+\ed x^2+x^2\ed u^2\,.
\end{align}
We treat the two photon rings at $r=\tilde{r}_\pm$ simultaneously.
First, we introduce a null frame
\begin{subequations}
\label{eq:NullFrame}
\begin{align}
	u^\mu\partial_\mu&=b\left(\partial_t+\tilde{\Omega}_\pm\partial_\theta\right)\,,\\
	v^\mu\partial_\mu&=-\left(1+\frac{\tilde{b}}{2}\right)\partial_t-\frac{b}{2}\tilde{\Omega}_\pm\partial_\theta\,,\\
	e^\mu\partial_\mu&=\frac{1}{\sqrt{D(\tilde{r}_\pm)}}\partial_r\,,
\end{align}
\end{subequations}
where $\tilde{\Omega}_\pm$ is given in Eq.~\eqref{eq:CriticalOmega}, $\gamma_{L\pm}$ is given in Eq.~\eqref{eq:LyapunovInTime}, and we introduced new quantities
\begin{align}
	b=\pm\frac{\sqrt{\mathcal{A}}}{\gamma_{L\pm}}\,,\qquad
    \mathcal{A}=3\left(\nu^2-1\right)\,,\qquad
    D(r)=\frac{1}{\left(\nu^2+3\right)(r-r_+)(r-r_-)}\,,
\end{align}
such that $u^\mu$ is a tangent vector to bound photon orbits, and on the photon ring
\begin{align}
	u^\mu u_\mu=v^\mu v_\mu=u^\mu e_\mu=v^\mu e_\mu=0\,\qquad
    u^\mu v_\mu=e^\mu e_\mu=1\,.
\end{align}
While this frame is adapted to bound photon orbits, it is not parallel-transported along them.
A parallel-transported vector satisfies the equations:
\begin{align}
\begin{split}
	\dot{V}^r(0)&=\mp\frac{\nu}{\sqrt{D(\tilde{r}_\pm)}}V^t(0)\pm\frac{\nu\tilde{\lambda}_\pm}{\sqrt{D(\tilde{r}_\pm)}}V^\theta(0)\,,\\
	V^r&=\dot{V}^r(0)\tau+V^r(0)\,,\\
	V^t&=2\nu\sqrt{D(\tilde{r}_\pm)}\left(\frac{1}{2}\dot{V}^r(0)\tau^2+V^r(0)\tau\right)+V^t(0)\,,\\
	V^\theta&=2\nu\lambda_\pm\sqrt{D(\tilde{r}_\pm)}\left(\frac{1}{2}\dot{V}^r(0)\tau^2+V^r(0)\tau\right)+V^\theta(0)\,,
\end{split}
\end{align}
where $\tau$ is an affine parameter and the dot represents the derivative with respect to this parameter.

A parallel-transported null frame attached to a bound photon orbit is then:
\begin{subequations}
\label{eq:PPFrame}
\begin{align}
	u^\mu\partial_\mu&=b\left(\partial_t+\tilde{\Omega}_\pm\partial_\theta\right)\,,\\
	v^\mu\partial_\mu&=-\partial_t+\left(\pm\tilde{\lambda}_\pm\nu^2\sqrt{D(\tilde{r}_\pm)}\tau^2-\frac{b}{2}\right)\left(\partial_t+\tilde{\Omega}_\pm\partial_\theta\right)\pm\frac{\nu\tau}{\sqrt{D(\tilde{r}_\pm)}}\partial_r\,,\\
	e^\mu\partial_\mu&=2\nu\tau\sqrt{D(\tilde{r}_\pm)}\left(\tilde{\lambda}_\pm\partial_t+\partial_\theta\right)+\frac{1}{\sqrt{D(\tilde{r}_\pm)}}\partial_r\,.
\end{align}
\end{subequations}
We define new coordinate $(u,v,x)$ such that $\partial_u=u^\mu\partial_\mu$, $\partial_v=v^\mu\partial_\mu$ and $\partial_x=e^\mu\partial_\mu$.
Solving Eq.~\eqref{eq:PPFrame}, we find that these coordinates are related to $(t,r,\theta)$ by the linear transformation
\begin{align}
\label{eq:FrameCoordinates}
    t=bu-\left(1+\frac{b}{2}\right)v\,,\qquad
	r=\tilde{r}_\pm\mp\frac{b}{2\tilde{\lambda}_\pm}x\,,\qquad
	\theta=\frac{b}{\tilde{\lambda}_\pm}\left(u-\frac{v}{2}\right)\,.
\end{align}
We now perform this change of coordinates in the original form of the metric \eqref{eq:WBTZ} and then rescale
\begin{align}
	(u,v,x)\rightarrow(u,k^2v,kx)\,,\qquad
    ds^2\rightarrow\frac{ds^2}{k^2}\,,
\end{align}
before taking the so-called Penrose limit $k\rightarrow 0$ to find
\begin{align}
	ds^2=2\ed u\ed v+\ed x^2+\mathcal{A}x^2\ed u^2+\mathcal{O}(k)\,,
\end{align}
where $u=\tau$ is the affine parameter along the bound null orbit, $v$ is constant along the wavefront (the set of all points that share the same phase), $x$ is the transverse coordinate associated to the direction $e^\mu$, and
\begin{align}
    \mathcal{A}=R_{\mu\alpha\nu\beta}u^\mu e^\alpha u^\nu e^\beta
    =3\left(\nu^2-1\right)
\end{align}
is the wave profile.
After a last change of coordinates to absorb the constant wave profile $A$,
\begin{align}
    u\rightarrow\frac{u}{\sqrt{A}}\,\qquad
    v\rightarrow\sqrt{\mathcal{A}}v\,,
\end{align}
we finally recover \eqref{eq:WarpedPenrose}.

\subsubsection{QNMs from the Penrose limit}

We have shown that the Penrose limit of the different photon rings takes the form \eqref{eq:WarpedPenrose}, and we have identified a coordinate transformation \eqref{eq:FrameCoordinates} linking Eq.~\eqref{eq:WarpedPenrose} to our initial spacetime metric \eqref{eq:WBTZ}.
Now, we investigate the quasinormal modes in the limiting Penrose geometry.

The mode solutions of the wave equation with outgoing boundary conditions in the background \eqref{eq:WarpedPenrose} were already obtained in \cite{Fransen:2023eqj}:
\begin{align}
    \label{eq:PenroseQNM}
	\Phi_n(u,v,x)\sim e^{-\left(n+\frac{1}{2}\right)u+ip_v\left(v+\frac{x^2}{2}\right)}H_n\left(-\sqrt{-ip_v}x\right)\,,
\end{align}
where $H_n(x)$ are Hermite functions.
We wish to compare them to the quasinormal modes \eqref{eq:ModeAnsatz} in the original spacetime.
After the change of coordinates \eqref{eq:FrameCoordinates}, the modes $e^{-i\omega t+ik\theta}\phi(r)$ have a phase that matches the phase of the modes \eqref{eq:PenroseQNM} under the identification
\begin{align}
	-i\frac{b}{\sqrt{A}}\omega+i\frac{b}{\sqrt{A}}\tilde{\Omega}_\pm k&=\mp\left(n+\frac{1}{2}\right)\,,\\
	\left(1+\frac{b}{2}\right)\omega-\frac{b}{2}\tilde{\Omega}_\pm k&=p_v\,.
\end{align}
Hence, the resonant modes in this limit must obey the same dispersion relation as in the last section:
\begin{align}
    \label{eq:PenroseSpectrum}
	\omega=\tilde{\Omega}_\pm k-i\left(n+\frac{1}{2}\right)\gamma_{L\pm}\,.
\end{align}
Thus, we see that the spectra \eqref{eq:EikonalCase1} and \eqref{eq:EikonalCase3} associated with the outer and inner photon rings, respectively, can be derived from the Penrose limit into their respective rings.
This conclusion is consistent with the findings in \cite{Fransen:2023eqj}, in which the eikonal QNM spectra of the Schwarzschild, Kerr, and self-dual WAdS$_3$ black holes were also obtained from the Penrose limits into their photon rings. 

\subsection{Emergent symmetry of the eikonal QNM spectrum}
\label{subsec:EikonalSymmetry}

In this section, we analyze the symmetries of the QNM spectrum within the near-ring region of phase space, and also solve the massless wave equation in that region.

We already defined the near-ring region in geodesic phase space in Eq.~\eqref{eq:NearRingRegion}.
Now we extend its definition to the space of scalar wave configurations.
To do so, we once again study the QNM potential \eqref{eq:WavePotential}.
In the eikonal limit where $\omega_R\sim k$ are both large and of comparable magnitude, the real part of the QNM potential matches the potential for null geodesics \eqref{eq:RadialPotential} if we identify
\begin{eqnarray}
    \omega_R\leftrightarrow E\,,\qquad
    k\leftrightarrow L\,.
\end{eqnarray}
We simultaneously describe the near-ring region for each of the two photon rings of the warped AdS$_3$ black hole, using the subscript $\pm$ to refer to their respective parameters.
By analogy with Eq.~\eqref{eq:NearRingRegion}, and following \cite{Hadar:2022xag,Kapec:2022dvc}, we define the near-ring region as
\begin{align}
    \label{eq:WaveNearRing}
    \text{NEAR-RING REGION:}\quad
    \begin{cases} 
        |\delta r|\ll\eta
        &\text{(near-peak)}\,,\\
        \left|\frac{k}{\omega_R}-\tilde{\lambda}_\pm\right|\ll\eta
        &\text{(near-critical)}\,,\\
        \frac{1}{\omega_R}\ll\eta
        &\text{(high-frequency)}\,,
    \end{cases}
\end{align}
where the radial deviation $\delta r=r-\tilde{r}_\pm$ was defined in Eq.~\eqref{eq:NearRingVariation} and the parameter $\eta=r_++r_-$ in Eq.~\eqref{eq:OrbitalRadii}.
The final condition requires eikonal modes to have a real part, which excludes the purely imaginary resonant modes from the discussion in this section.

Within this near-ring region, the QNM potential \eqref{eq:WavePotential} takes the form 
\begin{align}
    V_{\rm QNM}(\delta r)=3\left(\nu^2-1\right)\omega_R\delta r^2\mp2i\frac{\nu^2+3}{\sqrt{3\left(\nu^2-1\right)}}(r_+-r_-)\tilde{\lambda}_\pm\omega_R \omega_I\,.
\end{align}
Hence, the radial ODE \eqref{eq:RadialODE} can be rewritten as
\begin{align}
    \mathcal{H}\phi(\delta r)=i\omega_I\phi(\delta r)\,,
\end{align}
where we defined the Hamiltonian
\begin{align}
    \mathcal{H}=-\frac{1}{2p_1\omega_R}\left(\partial_{\delta r}^2+p_2^2\omega_R^2\delta r^2\right)\,,
\end{align}
with parameters
\begin{align}
    p_1=\frac{p_2}{\gamma_{L\pm}}\,,\qquad
    p_2=\frac{12\sqrt{3}\left(\nu^2-1\right)^{\frac{3}{2}}}{(r_+-r_-)^2\left(\nu^2+3\right)^2}\,.
\end{align}
This means that the massless wave equation in the near-ring region \eqref{eq:WaveNearRing} reduces (on scalar modes) to a time-independent Schrödinger equation corresponding to an inverted harmonic oscillator with non-Hermitian boundary conditions.
This explains the imaginary eigenvalues $i\omega_I$ and the appearance of the Hermite polynomials in Eq.~\eqref{eq:PenroseQNM}.
Following \cite{Subramanyan:2020fmx,Raffaelli:2021gzh,Hadar:2022xag}, we define the operators
\begin{align}
    a_\pm=\frac{e^{\pm\frac{p_2}{p_1}t}}{\sqrt{2p_2\omega_R}}\left(\mp i\partial_{\delta r}-p_2\omega_R\delta r\right)\,,\qquad
    L_0=-\frac{i}{4}(a_+a _-+a_-a_+)=\frac{ip_1}{2p_2}\mathcal{H}\,,\qquad
    L_\pm=\pm\frac{a_\pm^2}{2}\,.
\end{align}
The $a_\pm$ generate the Heisenberg algebra, and thus the $L_m$ satisfy the $\mathsf{SL}(2,\mathbb{R})_{\rm QN}$ algebra
\begin{align}
    \label{eq:Commutators}
    [a_+,a_-]=i\textbf{1}\,,\qquad
    [L_0,L_\pm]=\mp L_\pm\,,\qquad
    [L_+,L_-]=2L_0\,.
\end{align}
Note that for $a_\pm$ and $L_\pm$, the subscript $\pm$ no longer refers to the different photon rings---it is the standard notation for ladder operators.
In the definitions of these operators, the only dependence on the photon rings is encapsulated in the coefficient $p_1$, and the definitions remain the same for both $\tilde{r}_\pm$.
In principle, these operators are defined everywhere in our spacetime, but we focus on the near-ring region where $L_0$ is proportional to the Hamiltonian $\mathcal{H}$.
We also define the Casimir
\begin{align}
    \mathcal{C}=L_0^2-L_0-L_-L_+
    =L_0^2+L_0-L_+L_-
    =L_0^2-\frac{L_+L_-+L_-L_+}{2}\,,
\end{align}
which, by Schur's lemma, is proportional to the identity.
Here, the proportionality coefficient is
\begin{align}
    \mathcal{C}=-\frac{3}{16}\textbf{1}\,.
\end{align}
The eigenstates of $L_0$ satisfy the relation 
\begin{align}
    L_0\phi_h=h\phi_h\,,
\end{align} 
with eigenvalues $h$.
We thereby identify
\begin{align}
    \omega_I=-2\frac{p_2}{p_1}
    h=-2\gamma_{L\pm}h\,.
\end{align}
The mode ansatz \eqref{eq:ModeAnsatz} in the near-ring region \eqref{eq:WaveNearRing} reduces to 
\begin{align}
    \Phi(t,\delta r,\theta)=e^{-i\omega_Rt+ik\theta}\Phi_h(t,\delta r)\,,\qquad
    \Phi_h(t,\delta r)=e^{\omega_It}\phi_h(\delta r)
    =e^{-2\frac{p_2}{p_1}ht}\phi_h(\delta r)\,.
\end{align}

Henceforth, we restrict our attention to the outer photon ring, though the following computations and results are entirely identical for the inner photon ring.
Imposing outgoing boundary conditions for the fundamental modes is equivalent to imposing the highest-weight condition \cite{Kapec:2022dvc}
\begin{align}
    \label{eq:HighestWeight}
    L_+\Phi_h=0
    \qquad\iff\qquad
    a_+^2\Phi_h=0\,.
\end{align}
This requirement implies that
\begin{align}
    \mathcal{C}\Phi_h=h(h-1)\Phi_h
    \qquad\iff\qquad
    h=\frac{1}{4}\quad\text{or}\quad h=\frac{3}{4}\,.
\end{align}
As the highest-weight condition is a second-order differential equation, there exist two independent solutions $\Phi_{1,h}$ and $\Phi_{2,h}$, which obey
\begin{align}
    \label{eq:HighestWeightDetails}
    a_+\Phi_{1,h}=0
    \qquad\text{and}\qquad
    \Phi_{2,h}=a_-\Phi_{1,h}\,.
\end{align}
The fields defined by these conditions are manifestly independent and obey the highest-weight condition \eqref{eq:HighestWeight}.
Moreover, the commutation relations \eqref{eq:Commutators} imply that
\begin{subequations}
\begin{align}
    a_+\Phi_{2,h}&=i\Phi_{1,h}\,,\\
    a_+a_-\Phi_{1,h}&=i\Phi_{1,h}\,,\\
    a_-a_+\Phi_{2,h}&=i\Phi_{2,h}\,.
\end{align}  
\end{subequations}
In turn, these relations imply that
\begin{align}
    L_0\Phi_{1,h}=\frac{1}{4}\Phi_{1,h}\,,\qquad
    L_0\Phi_{2,h}=\frac{3}{4}\Phi_{2,h}\,.
\end{align}
As such, $\Phi_1$ has conformal weight $h=\frac{1}{4}$ and $\Phi_2$ has conformal weight $h=\frac{3}{4}$.
From Eq.~\eqref{eq:HighestWeightDetails}, we find their explicit forms:
\begin{align}
    \Phi_{1,h}=e^{-\frac{1}{2}\gamma_{L+}t+\frac{i}{2}p_2\omega_R\delta r^2}\,,\qquad
    \Phi_{2,h}=\delta re^{-\frac{3}{2}\gamma_{L+}t+\frac{i}{2}p_2\omega_R\delta r^2}\,.
\end{align}
The higher overtones are obtained as the towers of $\mathsf{SL}(2,\mathbb{R})_{\rm QN}$-descendants,
\begin{align}
    \Phi_{h, N}(t,\delta r)=L_-^N\Phi_h(t,\delta r)
    &=e^{-2\gamma_{L+}(h+N)t}\phi_{h+N}(\delta r)\\
    &\propto e^{-2\gamma_{L+}(h+N)t}D_{2(h+N)-\frac{1}{2}}\left(\sqrt{-2ip_2\omega_R}\delta r\right)\,,
\end{align}
where $D_n(\delta r)$ denotes the $n^\text{th}$ parabolic cylinder function.
With $n=2(h+N)-\frac{1}{2}$, near the edges $\delta r\to\pm\infty$ of the near-ring region, we have
\begin{align}
    \lim_{\delta r\to\pm\infty}D_n\left(\sqrt{-2ip_2\omega_R}\delta r\right)\approx\delta r^ne^{\frac{i}{2}p_2\omega_R\delta r^2}\,,
\end{align}
and the $n^\text{th}$ overtone near the edges is therefore
\begin{align}
    \Phi_n(t,\delta r,\theta)\approx e^{-\gamma_{L+}\left(n+\frac{1}{2}\right)t}\delta r^ne^{-i\omega_R\left(t-\frac{1}{2}p_2\delta r^2\right)+ik\theta}\,.
\end{align}
Reading off the frequencies of these overtones, we recover
from which we may read off the associated QNM spectrum to recover
\begin{align}
    \omega=\tilde{\Omega}_+k-i\left(n+\frac{1}{2}\right)\gamma_{L+}\,,
\end{align}
which is precisely the eikonal limit of the exact QNM spectrum \eqref{eq:EikonalCase1}, as expected.

The discussion is exactly identical for the inner photon ring and its associated eikonal spectrum.
Due to the exponential time dependence near the inner photon ring, we impose outgoing boundary conditions, which are equivalent to the highest-weight condition \eqref{eq:HighestWeight}, and the overtones are again obtained as $\mathsf{SL}(2,\mathbb{R})_{\rm QN}$-descendants
\begin{align}
    \Phi_{h,N}(t,\delta r)=L_-^N\Phi_h(t,\delta r)\,.
\end{align}
Near the edges, the $n^{\text{th}}$ overtone behaves like
\begin{align}
    \Phi_n(t,\delta r,\theta)\approx e^{-\gamma_{L-}\left(n+\frac{1}{2}\right)t}\delta r^n e^{-i\omega_R\left(t-\frac{1}{2}p_2\delta r^2\right)+ik\theta}\,,
\end{align}
from which we may read off the associated QNM spectrum to recover
\begin{align}
    \omega=\tilde{\Omega}_-k-i\left(n+\frac{1}{2}\right)\gamma_{L-}\,,
\end{align}
which is precisely the eikonal limit of the exact QNM spectrum \eqref{eq:EikonalCase3}, as expected.

In summary, we have shown that the eikonal QNM spectrum of the WAdS$_3$ black hole \eqref{eq:WBTZ} is controlled in the near-ring region by an emergent conformal symmetry $\mathsf{SL}(2,\mathbb R)_{\rm QN}$.
This conclusion is consistent with analogous findings for the Kerr black hole \cite{Hadar:2022xag} and self-dual WAdS$_3$ \cite{Kapec:2022dvc}, and of course with the Penrose-limit spectrum \eqref{eq:PenroseSpectrum}.

\subsection{QNMs from geometric optics}
\label{subsec:GeometricOptics}

The final method we will apply to the warped AdS$_3$ black hole \eqref{eq:WBTZ} to derive its (eikonal) QNMs and their frequency spectrum is the geometric-optics approximation.
This method has been primarily used for Schwarzschild and Kerr black holes, for which the exact QNM spectrum does not admit a simple analytic representation (though recent work has derived a representation in terms of the Nekrasov partition function \cite{Bonelli:2021uvf}).
Despite providing only approximate results, this method has the advantage of having a clear physical interpretation, and it can be used to check whether the exact spectrum obtained from our boundary conditions behaves as expected.

In geometric optics, solutions to the massless wave equation in the high-frequency limit are approximated by congruences of null geodesics, with collections of particles flowing along these geodesics in phase describing wavefronts of the field.
As reviewed in Sec.~\ref{subsec:Geometrization}, the QNMs of a Kerr black hole are precisely described in this limit by null geodesics that asymptote to bound photon orbits around the black hole.
Here, we show that the same is true for warped AdS$_3$ black holes.

When the wave frequencies are large compared to the local curvature, solutions to the massless wave equation \eqref{eq:Laplacian} take the approximate form
\begin{align}
    \Phi\approx A\,e^{iS}\,,
\end{align}
where $S(x^\mu)$ is a rapidly oscillating phase and $A(x^\mu)$ a slowly varying amplitude. 
The wave equation is then expressed in terms of the gradient of the phase
\begin{align}
    p_\mu = \partial_\mu S \, ,
\end{align}
as
\begin{align}
    \label{eq:GeometricWaves}
    -p_\mu p^\mu A+i\left(2p^\mu\nabla_\mu A+\nabla_\mu p^\mu A\right)+\nabla^2A=0\,.
\end{align}
In the geometric-optics approximation, one solves this equation order-by-order in $p_\mu$.
At leading order, it implies that $p_\mu$ is a null vector
\begin{align}
    p_\mu p^\mu=0\,.
\end{align}
This implies that the phase $S(x^\mu)$ is a solution to the Hamilton-Jacobi equation.
Thus, $p_\mu$ must satisfy the geodesic equation
\begin{align}
    p^\mu\nabla_\mu p_\nu=0\,,
\end{align}
and it naturally defines an affine parameter $s$ along geodesics via
\begin{align}
    \partial_s=p^\mu \partial_\mu\,.
\end{align}
At subleading order, Eq.~\eqref{eq:GeometricWaves} relates the expansion $\hat{\theta}=\nabla_\mu p^\mu$ to the parallel transport of the amplitude via
\begin{align}
    \label{eq:SubleadingWave}
    p^\mu\nabla_\mu A=-\frac{1}{2}\hat{\theta}A\,.
\end{align}

As we have seen, the wave equation and geodesic motion in the warped AdS$_3$ black hole are both separable.
The general solution of the Hamilton-Jacobi equation separates into
\begin{align}
    S=-\omega t+k\theta+\kappa(r)\,,
\end{align}
where the function $\kappa(r)$ satisfies the equation
\begin{align}
    \kappa'(r)^2=g_{rr}^2+V_{\rm QNM}(r)\,,
\end{align}
with potential $V_{\rm QNM}(r)$ given in Eq.~\eqref{eq:WavePotential}.
For null geodesics that asymptote to the (outer) photon ring with $\frac{k}{\omega}=\tilde{\lambda}_+$,
\begin{align}
    \label{eq:RadialBehavior}
    \kappa(r)&=\pm\omega\frac{\sqrt{3\left(\nu^2-1\right)}}{\nu^2+3}\int\frac{r-\tilde{r}_+}{(r-r_+)(r-r_-)}\ed r\\
    &=\pm\frac{\omega\nu}{\nu^2+3}\left(\ln\left[\frac{r-r_-}{r-r_+}\right]+\frac{\sqrt{3\left(\nu^2-1\right)}}{2\nu}\ln\left[(r-r_+)(r-r_-)\right]\right)\,.
\end{align}
In the near-ring region, this function may be approximated (up to an irrelevant constant) as
\begin{align}
    \label{eq:NearRingBehavior}
    \kappa(r)\approx\pm\frac{6\left(\nu^2-1\right)\omega}{\left(\nu^2+3\right)(r_+-r_-)^2}\delta r^2\,,
\end{align}
which is consistent with the behavior found in the Penrose limit and in the near-ring region approaches.
Since we are interested in geodesics that leak out of the photon ring, we choose the positive sign in Eq.~\eqref{eq:NearRingBehavior} and consequently in Eq.~\eqref{eq:RadialBehavior}.

The expansion of this congruence is 
\begin{align}
    \hat{\theta}=\sqrt{3\left(\nu^2-1\right)}\omega\,,
\end{align}
and is constant as mentioned earlier.
We now go on to tackle the subleading order equation \eqref{eq:SubleadingWave}.
Its solution is defined in the whole spacetime, but we will restrict ourselves to the near-ring region, where it takes a more elegant form that allows for comparison with the various previous approaches:
\begin{align}
    \label{eq:ApproximateA}
    \left(\delta r\,\partial_{\delta r}+\frac{1}{\gamma_{L+}}\partial_t+\frac{1}{2}\right)A(t,\delta r)=0\,.
\end{align}
The general solution to this equation is
\begin{align}
    A(t,\delta r)=\mathcal{A}\left(e^{-\gamma_{L+}t}\delta r\right)e^{-\frac{\gamma_{L+}}{2}t}\,,
\end{align}
where $\mathcal{A}$ can be any function that is regular at $\delta r=0$.
Such a function may be series-expanded around the photon ring in $\delta r=0$, such that the general solution to \eqref{eq:ApproximateA} reads
\begin{align}
    A(t,\delta r)=\sum_n\mathcal{A}_n\delta r^ne^{-\left(n+\frac{1}{2}\right)\gamma_{L+} t}\,.
\end{align}
The general solution in the geometric-optics approximation in the near-ring region is
\begin{align}
    \Phi_n^{\rm eikon}\approx A_n(t,\delta r)e^{iS}
    =\delta r^n\exp\left(-i\omega^{\rm eikon}t+k\theta+\frac{6\left(\nu^2-1\right)\tilde{\Omega}_+k}{\left(\nu^2+3\right)(r_+-r_-)^2}\delta r^2\right)\,,
\end{align}
with
\begin{align}
    \omega^{\rm eikon}=\tilde{\Omega}_+k-i\left(n+\frac{1}{2}\right)\gamma_{L+}\,.
\end{align}
From these results, we see that the exact resonant modes \eqref{eq:QNMsCase1} with eikonal limit \eqref{eq:EikonalCase1}, which were defined by imposing the usual QNM boundary conditions (purely ingoing at the horizon and purely outgoing at infinity) in the warped AdS$_3$ background \eqref{eq:WBTZ} are sourced by the outer photon ring, which may be viewed as their geometrization in the same way as the Kerr photon shell geometrizes the eikonal Kerr QNM spectrum.

\subsection{Conformal symmetry of the warped photon ring}
\label{subsec:ConformalSymmetry}

In this section, we consider the observational appearance of the WAdS$_3$ black hole as seen by a distant observer in this three-dimensional spacetime.
We are particularly interested in its critical curve, which is the image of its (outer) photon ring and whose Kerr analogue was reviewed in Sec.~\ref{subsec:CriticalCurve}.
In $(2+1)$-dimensions, the observer screen is simply a line, and the black hole photon ring is a single critical point at infinity rather than a closed critical curve as in higher dimensions.

Light rays traveling close to the outer photon ring can undergo several rotations around the black hole before escaping to reach the distant observer.
This implies that a single source can produce multiple images, indexed by the azimuthal winding number of the photons that produce them.
Here, we construct an observable symmetry group $\mathsf{SL}(2,\mathbb{R})_{\rm PR}$ of the photon ring, as was done for the Schwarzschild and Kerr black holes \cite{Hadar:2022xag} and self-dual warped AdS$_3$ \cite{Kapec:2022dvc}.
The dilation generator of this conformal symmetry relates successive images of the photon ring with increasing photon winding number.

Let $\tilde{\Gamma}=\{(t,r,\theta,p_t,p_r,p_\theta)\}$ and $\Gamma=\{(r,\theta,p_r,p_\theta)\}$ denote the extended and reduced phase spaces of null geodesics in the geometry \eqref{eq:WBTZ}, and let $\Omega$ be the canonical symplectic form on $\Gamma$,\footnote{The procedure for reducing the extended phase space $\tilde{\Gamma}$ to $\Gamma$ is discussed in the Appendix of \cite{Kapec:2022dvc}.}
\begin{align}
    \ed\Omega=\ed p_r\wedge\ed r+\ed p_\theta\wedge\ed\theta\,.
\end{align}
The Hamiltonian is obtained by solving the null condition $g^{\mu\nu}p_\mu p_\nu=0$ for $p_t=-H$, which gives
\begin{align}
    \label{eq:GeodesicH}
    H=-\frac{m(r)}{n(r)}p_\theta+\frac{1}{2n(r)D(r)}\sqrt{D(r)p_\theta^2+n(r)p_r^2}\,,
\end{align}
where we introduced the functions
\begin{align}
    D(r)=\frac{1}{4R(r)^2N(r)^2}\,\qquad
    n(r)=R(r)^2\,\qquad
    m(r)=R(r)^2N^\theta(r)\,.
\end{align}
The Hamiltonian \eqref{eq:GeodesicH} reduces on the (outer) photon ring $(r,p_r)=(\tilde{r}_+,0)$ to the critical energy $\tilde{H}=\frac{p_\theta}{\tilde{\lambda}_+}$.
Since this dynamical system is integrable, it admits a canonical transformation to action-angle variables $(r,p_r,\theta,p_\theta)\to(T,H,\Theta,L)$ that preserves the symplectic form $\ed\Omega$:
\begin{subequations}
\begin{align}
    \ed T&=\frac{2D(r)}{\sqrt{\mathcal{V}(r)}}\left[n(r)H+m(r)L\right]\ed r\,,\\ 
    \ed\Theta&=\ed\theta+\frac{2D(r)}{\sqrt{\mathcal{V}(r)}}\left[m(r)H+L\right]\ed r\,,\\ 
    L&=p_\theta\,,
\end{align}
\end{subequations}
where $\mathcal{V}(r)$ is related to the radial geodesic potential \eqref{eq:RadialPotential} by a simple constant,
\begin{align}
    \label{eq:V}
    \mathcal{V}(r)\equiv\frac{p_r^2}{4D(r)^2}
    =\left[L^2+2m(r)HL+n(r)H^2\right]\,.
\end{align}
In these variables, the equations of motion are simply
\begin{subequations}
\label{eq:Canonical}
\begin{align}
    \dot{H}=\{H,H\}=0\,,
    &\qquad
    \dot{L}=\{L,H\}=0\,,\\ 
    \dot{\Theta}=\{\Theta,H\}=0\,,
    &\qquad
    \dot{T}=\{T,H\}=1\,.
\end{align}
\end{subequations}
The first equation on the second line implies that a photon with initial coordinates $(r_i,\theta_i,H,L)$ evolves to final coordinates $(r_f,\theta_f,H,L)$ according to:
\begin{align}
    \Delta\theta\equiv\theta_f-\theta_i
    =\int_{\theta_i}^{\theta_f}\ed\theta
    =-2\int_{r_i}^{r_f}\frac{D(r)}{\sqrt{\mathcal{V}(r)}}\left[m(r)H+L\right]\ed r\,.
\end{align}
The final equation in Eqs.~\eqref{eq:Canonical} implies that the time elapsed along such a trajectory is:
\begin{align}
    T=2\int_{r_i}^{r_f}\frac{D(r)}{\sqrt{\mathcal{V}(r)}}\left[n(r)H+m(r)L\right]\ed r\,.
\end{align}
Since we are focusing on optical images, we will only consider geodesics that begin and end at infinity, always remaining outside the ring orbit at $r=\tilde{r}_+$.
An observer at infinity receives these null geodesics with impact parameter
\begin{align}
    \lambda=\frac{L}{H}>\tilde{\lambda}_+\,,
\end{align}
and energy
\begin{align}
   \hat{H}=H-\frac{L}{\tilde{\lambda}_+}\,.
\end{align}
The radius of closest approach $r_{\rm min}$ of a geodesic with impact parameter $\lambda$ is reached when its radial momentum $p_r$ vanishes, which by Eq.~\eqref{eq:V}, is equivalent to requiring that $\mathcal{V}(r_{\rm min})=0$.
Solving this condition yields
\begin{align}
    \label{eq:Perihelion}
    r_{\rm min}=\tilde{r}_+-\frac{2}{3\left(\nu^2-1\right)}\left[2\nu\left(\lambda-\tilde{\lambda}_+\right)-\sqrt{\left(\nu^2+3\right)\left(\lambda-\tilde{\lambda}_+\right)\left(\lambda-\tilde{\lambda}_-\right)}\right]\,.
\end{align}
Geodesics with $\hat{H}=0$ are asymptotic to the photon orbit at $r=\tilde{r}+$ in the far past and/or future.
Their impact parameter $\lambda=\tilde{\lambda}_+$ defines the critical point on the observed line at infinity.
On the geodesic phase space, we may define the action of the group $\mathsf{SL}(2,\mathbb{R})_{\rm PR}$ with generators
\begin{align}
    H_+=\hat{H}\,,\qquad
    H_0=-T\hat{H}\,,\qquad
    H_-=T^2\hat{H}\,.
\end{align}
This algebra commutes with the $\mathsf{U}(1)$ algebra generated by $L$ and thus preserves the hypersurfaces $\Gamma_L$ of fixed angular momentum.
However, it does not preserve constant-energy hypersurfaces and therefore modifies the impact parameter and hence the radius of closest approach of geodesics.

Next, we construct a discrete dilation operator that acts on the phase space of observed null geodesics and increasing the winding number of null geodesics to relate successive images of a source.
The photon ring is an attractive fixed point for the flow generated by the one parameter dilation operator $e^{-\alpha H_0}$ \cite{Hadar:2022xag, Kapec:2022dvc}, under which 
\begin{align}
    \label{eq:Dilation}
    \hat{H}(0)\to\hat{H}(\alpha)=e^{-\alpha}\hat{H}(0)\,,
\end{align}
because $\hat{H}(\alpha)$ satisfies
\begin{align}
    \partial_\alpha\hat{H}(\alpha)=\{H_0,\hat{H}(\alpha)\}
    =-\hat{H}(\alpha)\,.
\end{align}
For large $\alpha$, $\hat{H}(\alpha)$ becomes small but $T(\alpha)\to\infty$.
Defining the dimensionless parameter $\delta r=\frac{r-\tilde{r}_+}{\tilde{r}_+}$, the point of closest approach \eqref{eq:Perihelion} becomes at leading order, for $\hat{H}(\alpha)\to0$,
\begin{align}
    \delta r_{\rm min}^2=\frac{4\left(\nu^2+3\right)}{9\left(\nu^2-1\right)^2}\frac{\tilde{\lambda}_+^2}{\tilde{r}_+^2}\left(\tilde{\lambda}_+-\tilde{\lambda}_-\right)\frac{\hat{H}(\alpha)}{L}\,,
\end{align}
while
\begin{align}
    \ed T(\alpha)\approx\frac{1}{\gamma_{L+}}\ed\ln{\delta r}
    \qquad\Longrightarrow\qquad
    \delta r\approx\delta r_0e^{\gamma_{L+}T(\alpha)}\,,
\end{align}
as expected.
It follows that under the $\mathsf{SL}(2,\mathbb{R})_{\rm PR}$ dilation \eqref{eq:Dilation},
\begin{align}
    \partial_\alpha\ln{\delta r_{\rm min}}=-\frac{1}{2}\,.
\end{align}
For a geodesic beginning and ending at $r=\infty$,
\begin{align}
    \Delta\theta=-4\int_{r_{\rm min}}^\infty\frac{D(r)}{\sqrt{\mathcal{V}(r)}}\left[m(r)H+L\right]\ed r
    \approx\frac{2\pi}{\gamma_{L+}}\ln{\delta r_{\rm min}}\,,
\end{align}
to leading order as $\delta r_{\rm min}\to 0$.
Then, the geodesic winding number $w=|\Delta\theta|/2\pi$ around the black hole diverges like
\begin{align}
    \partial_\alpha w=\frac{1}{2\gamma_{L+}}\,,
\end{align}
under dilations.
If we consider a source at $(r_s,\theta_s)$ and an ``observer'' at $(r_o,\theta_o)$, then there are an infinite number of null geodesics connecting them, labeled by their winding number $w$ and with the same angular shift $\Delta\theta$ modulo $2\pi$.
We conclude that for large $w$ (or equivalently, for small $\hat{H}$ or small $\delta r_{\rm min}$), if $\alpha=2\gamma_{L+}$, then the dilation
\begin{align}
    D_0=e^{-2\gamma_{L+}H_0}\,
\end{align}
maps a geodesic corresponding to an image of the source to its successive image, increasing the winding number $w\to w+1$.
The products of $D_0$ form a discrete subgroup of $\mathsf{SL}(2,\mathbb{R})_{\rm PR}$ mapping the phase space of observed null geodesics to itself.

\subsection{Extremal limits}
\label{subsec:ExtremalLimits}

In the extremal limit, the warped AdS$_3$ black hole develops a near-horizon throat, much like the Kerr black hole (Sec.~\ref{subsec:ExtremeKerr}).
In this section, we compute QNMs for extreme and near-extreme WAdS$_3$ black holes in the near-horizon limit, in which the geometry corresponds to self-dual warped AdS$_3$.
Our results for the extreme WAdS$_3$ black hole reproduce those of \cite{Kapec:2022dvc}.

\subsubsection{Near-horizon QNMs of extreme warped black holes}

First, we compute the QNMs in the extremal case where the two horizons coincide, $r_+=r_-=r_0$.
The metric \eqref{eq:WBTZ} then takes the form
\begin{align}
    ds^2&=\ed t^2+\frac{\ed r^2}{\left(\nu^2+3\right)(r-r_0)^2}+\left(2\nu r-r_0\sqrt{\nu^2+3}\right)\ed t\ed\theta\\
    &\phantom{=}+\frac{r^2}{4}\left[3\left(\nu^2-1\right)-\frac{2r_0}{r}\sqrt{\nu^2+3}\left(2\nu-\sqrt{\nu^2+3}\right)\right]\ed\theta^2\,.
\end{align}
A similar computation to the one in Sec.~\ref{subsec:PhotonRing} yields a unique photon ring at orbital radius
\begin{align}
    \tilde{r}=r_0\,,\qquad
    \tilde{\lambda}=-\frac{r_0}{2}\left(2\nu-\sqrt{\nu^2+3}\right)
    =\frac{1}{\tilde{\Omega}}\,,
\end{align}
which coincides with the horizon.
This makes it impossible to define a Lyapunov exponent as usual.

We now repeat the analysis of Sec.~\ref{subsec:WarpedQNMs} in this new setting.
In the extremal case, the new radial coordinate needed to recover a hypergeometric differential equation is
\begin{align}
    z=\frac{1}{r-r_0}\,,
\end{align}
in terms of which the horizon is now located at $z\to\infty$, while radial infinity is at $z\to0$, and $z$ is always positive within this range.

The radial part of the general mode solution to the massless scalar wave equation is
\begin{align}
    \phi(z)&=Q_-e^{-Az}z^C\,{_1F_1}(B+C;2C;2Az)+Q_+e^{-Az}z^{1-C}\,{_1F_1}(1+B-C;2-2C;2Az)\,,
\end{align}
where ${_1F_1}(a,b,x)$ is the confluent hypergeometric function, and its coefficients $A$, $B$, and $C$ are
\begin{subequations}
\begin{align}
    A&=\frac{i}{\nu^2+3}\left[2k+r_0\left(2\nu-\sqrt{\nu^2+3}\right)\omega\right]\,,\\
    B&=\frac{2i\nu\omega}{\nu^2+3}\,,\\
    C&=\frac{1}{2}\left(1-\sqrt{1-\frac{12\left(\nu^2-1\right)\omega^2}{(\nu^2+3)^2}}\right)\,.
\end{align}
\end{subequations}
Next, we must impose QNM boundary conditions, as we did in the sub-extremal case: ingoing at the horizon ($z\to\infty$) and outgoing at infinity ($z\to0$).
We begin with the second condition.
Near $z=0$, the radial part of the modes behaves as
\begin{align} 
    \label{eq:HorizonExtremeQNM}
    \phi(z)\stackrel{z\rightarrow0}{\approx}Q_+z^{\frac{1}{2}+\varpi}+Q_-z^{\frac{1}{2}-\varpi}\,.
\end{align}
where $\varpi=\frac{1}{2}-C$, as before.
In these coordinates, an ingoing wave has decreasing $z$, so we impose
\begin{align}
    \label{eq:InfinityCondition}
    Q_+=0.
\end{align}
Meanwhile, near the horizon at $z\to\infty$, the modes behave as
\begin{align}
    \phi(z)\stackrel{z\to\infty}{\approx}C_+e^{Az}z^B+C_-e^{-Az}z^{-B}\,,
\end{align}
with near-horizon coefficients $(C_+,C_-)$ related to the $(Q_+,Q_-)$ by the connection formulas
\begin{align}
    C_+=\frac{Q_+}{\Gamma(1+B-C)}+\frac{Q_-}{\Gamma(B+C)}\,,\qquad
    C_-=\frac{Q_+}{\Gamma(1-B-C)}+\frac{Q_-}{\Gamma(-B+C)}\,.
\end{align}
The wave is ingoing at the horizon if it moves towards positive values of $z$.
This selects
\begin{align}
    C_-=0\,.
\end{align}
Combined with the outgoing condition at infinity \eqref{eq:InfinityCondition}, this implies that the QNM condition is
\begin{align}
        -B+C=-n\quad\iff\quad-\frac{2i\nu}{\nu^2+3}\omega +\frac{1}{2}\left(1-\sqrt{1-\frac{12\left(\nu^2-1\right)}{\left(\nu^2+3\right)^2}\omega^2}\right)=-n\,.
\end{align}
This is exactly the condition \eqref{eq:Case2}, which leads to the spectrum of frequencies \eqref{eq:QNMsCase2},
\begin{align} 
    \omega_{2+}=-i\nu(2n+1)+i\sqrt{3\left(\nu^2-1\right)n(n+1)+\nu^2}\,,
\end{align}
which has no real part in the dispersion relation.
It would now appear that the usual eikonal QNM spectrum \eqref{eq:QNMsCase1} associated with the (outer) photon ring has disappeared, but it is only missing from this analysis because it implicitly assumes that $A\neq0$.
However, these modes have $\frac{k}{\omega_R}=\tilde{\lambda}$, which means that in the eikonal limit $A\to0$.
As such, we must separately consider the special case $A=0$.
In that case, the radial ODE that we need to solve simplifies to
\begin{align}
    \partial_z^2\phi(z)=\frac{4\varpi^2-1}{4z^2}\phi(z)\,,
\end{align}
with general solution
\begin{align}
    \phi(z)=Q_+z^{\frac{1}{2}+\varpi}+Q_-z^{\frac{1}{2}-\varpi}\,.
\end{align}
This solution looks similar to the one in Eq.~\eqref{eq:HorizonExtremeQNM}, but in this case, it is actually valid for any $z$, not just at spatial infinity.
Since we are interested in QNMs leaking out of the photon ring, and since the photon ring is precisely located at the horizon, it suffices in this case to require that the modes propagate towards spatial infinity (from $z=\infty$ to $z=0$), while the horizon condition is automatically satisfied.
In other words, we only need to impose one condition, Eq.~\eqref{eq:InfinityCondition}.
The dispersion relation for the resulting QNMs then recovers the spectrum \eqref{eq:QNMsCase1},  as expected:
\begin{align}
    \omega=\tilde{\Omega}k\,,
\end{align}
One can also obtain this solution from sending $r_+\to r_-=r_0$ in Eq.~\eqref{eq:QNMsCase1}.

\subsubsection{Near-horizon QNMs of near-extreme warped black holes}

Now, we want to take the near-horizon limit of the near-extreme warped black hole.
The resulting metric is the self-dual WAdS$_3$ spacetime whose photon ring and eikonal QNM spectrum were studied in \cite{Kapec:2022dvc}.
To take this limit, we first need the Hawking temperature at the outer horizon,
\begin{align}
    T_H=\frac{\nu^2+3}{4\pi}\frac{r_+-r_-}{2\nu r_+-\sqrt{r_+r_-\left(\nu^2+3\right)}}\,.
\end{align}
We also define a dimensionless, rescaled temperature
\begin{align}
    T_R=\frac{2\nu-\sqrt{\nu^2+3}}{\nu^2+3}\frac{T_H}{\varepsilon}\,,
\end{align}
where $\varepsilon>0$ is for now just a dummy parameter, such that the horizons are located at
\begin{align}
        r_\pm=r_0\left(1+2\pi T_R\varepsilon\right).
\end{align}

We also perform a coordinate transformation $(t,r,\theta)\to(\hat{t},\hat{\theta},\hat{r})$ to new coordinates
\begin{subequations}
\label{eq:NearExtremeCoordinates}
\begin{align} 
    \hat{t}&=-\frac{\nu^2+3}{2\nu-\sqrt{\nu^2+3}}\varepsilon t\,,\\
    \hat{\theta}&=\frac{\left(2\nu-\sqrt{\nu^2+3}\right)\left(\nu^2+3\right)}{4\nu}r_0\theta+\frac{\nu^2+3}{2\nu}t\,,\\
    \hat{r}&=\frac{r-r_0}{r_0\varepsilon}-2\pi T_R\,.
\end{align}
\end{subequations}
We now want to dial the black hole \eqref{eq:WBTZ} towards extremality while zooming into its horizon at the same rate, so as to maintain a small temperature (or deviation from extremality) within the near-horizon region.
That is, we scale $\varepsilon\to0$ and thus $T_H\to0$ while keeping $T_R$ fixed.
In this limit, the metric \eqref{eq:WBTZ} simplifies to
\begin{align} 
    \label{eq:SelfDual}
    ds^2=\frac{1}{\nu^2+3}\left(-\hat{r}\left(\hat{r}+4\pi T_R\right)\ed\hat{t}^2+\frac{\ed\hat{r}^2}{\hat{r}\left(\hat{r}+4\pi T_R\right)}+\Lambda^2\left[\ed\hat{\theta}+\left(\hat{r}+2\pi T_R\right)\ed\hat{t}\right]^2\right)\,,
\end{align}
with
\begin{align}
    \Lambda=\frac{2\nu}{\sqrt{\nu^2+3}}\,.
\end{align}
We recognize this as the metric \eqref{eq:NearNHEK} induced on polar slices of constant $\theta$ within the near-NHEK region of a near-extreme Kerr black hole (up to a radial shift $R\to R+\kappa$ with $\kappa=4\pi T_R$), except that in this case, the warp factor $\Lambda$ is set not by the choice of polar slice $\theta$, but rather by the black hole parameter $\nu$.
We emphasize that this scaling limit resolves the entire near-extremal throat region and is therefore different from the Penrose limit, which zooms into a single geodesic.

Under the change of coordinates \eqref{eq:NearExtremeCoordinates}, the photon ring radii transform to
\begin{align}
    \hat{r}_\pm=\frac{\tilde{r}_\pm-r_0}{r_0\varepsilon}-2\pi T_R
    =2\pi T_R\left(\pm\frac{\Lambda}{\sqrt{\Lambda^2-1}}-1\right)\,,
\end{align}
and the angular velocities and Lyapunov exponents of (nearly) bound orbits become
\begin{align}
        \hat{\Omega}_\pm=\mp2\pi T_R\sqrt{1-\frac{1}{\Lambda^2}}\,,\qquad\hat{\gamma}_{L\pm}=2\pi T_R\,.
\end{align} 
in agreement with \cite{Kapec:2022dvc}.

As we did for the full WAdS$_3$ black hole \eqref{eq:WBTZ}, one can also compute the exact QNM spectrum in the self-dual WAdS$_3$ geometry \eqref{eq:SelfDual}.
It was argued in \cite{Kapec:2022dvc} that its eikonal spectrum is
\begin{align} 
    \label{eq:RecallSpectrum}
    \hat{\omega}_\pm\stackrel{|k|\gg1}{\approx}\hat{\Omega}_\pm\hat{k}-i\left(n+\frac{1}{2}\right)\hat{\gamma}_{L\pm}\,.
\end{align}
From the perspective of our analysis of the full WAdS$_3$ black hole, this conclusion may seem surprising.
Indeed, we found that there were two reasonable but inequivalent choices of boundary conditions that one could impose at infinity, and that these led to the two different spectra \eqref{eq:OuterQNMs} and \eqref{eq:InnerQNMs}.
When taking the near-extreme, near-horizon limit of either one of these spectra, we are left with a single branch of the spectrum \eqref{eq:RecallSpectrum}.

For instance, suppose that we impose the outgoing condition at infinity in the full WAdS$_3$ black hole.
This results in the spectrum \eqref{eq:OuterQNMs} with two branches labeled 1 and 2.
In the near-extreme, near-horizon limit, the first branch survives and takes the expected form in the eikonal limit
\begin{align}
    \hat{\omega}_1\approx\hat{\Omega}_+\hat{k}-i\left(n+\frac{1}{2}\right)\hat{\gamma}_{L+}\,.
\end{align}
By contrast, the purely imaginary frequencies in the second branch diverge in this limit,
\begin{align}
    \hat{\omega}_{2}\sim\mathcal{O}\left(\frac{1}{\varepsilon}\right)\,,
\end{align}
and would therefore seem to disappear from the QNM spectrum associated with the near-horizon geometry.
Thus, for this choice of boundary condition in the full WAdS$_3$ black hole spacetime, we are seemingly left with only half of the expected eikonal spectrum \eqref{eq:RecallSpectrum} in the near-horizon geometry (in this case, its $+$ branch, while the $-$ branch would be recovered from the finite-flux boundary condition).

However, this does not contradict the findings of \cite{Kapec:2022dvc}, because if one imposes ``outgoing'' boundary conditions at spatial infinity \textit{within} the near-extreme, near-horizon geometry \eqref{eq:SelfDual}, then one still obtains \textit{both} branches of the eikonal spectrum \eqref{eq:RecallSpectrum}.
The reason is that the QNM frequencies \eqref{eq:InnerQNMs}, which in the full spacetime are excluded by the outgoing boundary condition at infinity ($r\to\infty$ in Eq.~\eqref{eq:WBTZ}), are associated in the geometry \eqref{eq:SelfDual} with modes that do satisfy its outgoing boundary condition.
As a result, they become admissible under this boundary condition and reproduce the missing ($-$) branch of \eqref{eq:RecallSpectrum}.

\section*{Acknowledgements}

The authors thank Shahar Hadar, Kristiansen Lara, and Abhishek Pathak for useful discussions and exchanges on the topics covered in this work.
SD is a Senior Research Associate of the Fonds de la Recherche Scientifique F.R.S.-FNRS (Belgium).
SD and RW acknowledge support of the Fonds de la Recherche Scientifique F.R.S.-FNRS (Belgium) through the projects PDR/OL C62/5 ``Black hole horizons: away from conformality'' (2022-2025) and CDR n°40028632 (2025-2026). 
This work is supported by the F.R.S.-FNRS (Belgium) through convention IISN 4.4514.08 and benefited from the support of the Solvay Family. SD and QV are members of BLU-ULB, the interfaculty research group focusing on space research at ULB - Université libre de Bruxelles.
AL is supported by the National Science Foundation under grants AST-2307887 and PHY-2340457.
RW also acknowledges support by the Heising-Simons Foundation under the ``Observational Signatures of Quantum Gravity'' collaboration grant 2021-2818 and the U.S. Department of Energy, Office of High Energy Physics,
under Award No. DE-SC0019470.

\appendix

\section{Warped black holes in the quadratic ensemble}
\label{app:QuadraticEnsemble}

In this appendix, we give expressions for the QNM frequencies of the WAdS$_3$ black hole \eqref{eq:WBTZ} in another commonly used set of ``quadratic ensemble'' coordinates  \cite{Detournay:2012pc,Detournay:2015ysa}.
The warping parameter in this case is denoted by ``$H$'' and such that when $H=0$, one recovers the BTZ black hole in the usual coordinates.
This will enable us to compare our results to the well-known BTZ QNMs \cite{Birmingham:2001hc}:
\begin{align} 
    \label{eq:SpectrumBTZ}
    \hat{\omega}_L=\hat{k}-2i\left(\hat{r}_+-\hat{r}_-\right)\left(n+1\right)\,,\qquad
    \hat{\omega}_R=-\hat{k}-2i\left(\hat{r}_++\hat{r}_-\right)\left(n+1\right)\,.
\end{align}
Our starting point is the metric \eqref{eq:WBTZ}, for which the change of coordinates to the quadratic ensemble is well known \cite{Aggarwal:2023peg}.
In the quadratic ensemble, this black hole has the ADM-form line element
\begin{align}
    \label{eq:QE}
    ds^2=-N_{\QE}(\hat{r})^2\ed\hat{t}^2+\frac{\left(1-2H^2\right)\hat{r}^2}{R_{\QE}(\hat{r})^2N_{\QE}(\hat{r})^2}\ed\hat{r}^2+R_{\QE}(\hat{r})^2\left[\ed\hat{\theta}-N^{\hat{\theta}}(\hat{r})\ed\hat{t}\right]^2\,, 
\end{align}
which depends on two parameters $H$ and $L$ and the functions
\begin{subequations}
\begin{align}
    R_{\QE}(\hat{r})^2&=\hat{r}^2-2H^2\left(\frac{\hat{r}^2+\hat{r}_+\hat{r}_-}{\hat{r}_++\hat{r}_-}\right)^2\,,\\
    N_{\QE}(\hat{r})^2&=\frac{(1-2H^2)}{R_{\QE}(\hat{r})^2L^2}\left(\hat{r}^2-\hat{r}_+^2\right)\left(\hat{r}^2-\hat{r}_-^2\right)\,,\\
    N^{\hat{\theta}}(\hat{r})&=\frac{1}{R_{\QE}(\hat{r})^2L}\left[\left(1-2H^2\right)\hat{r}+2H^2\frac{\left(\hat{r}^2-\hat{r}_+^2\right)\left(\hat{r}^2-\hat{r}_-^2\right)}{\left(\hat{r}_++\hat{r}_-\right)}\right]\,,
\end{align}
\end{subequations}
where the subscript ``QE'' (quadratic ensemble) is meant to differentiate these functions from those in Eq.~\eqref{eq:WBTZ}.
This metric has the nice feature of reducing to the usual BTZ metric when $H\to0$.
The line elements \eqref{eq:QE} are \eqref{eq:WBTZ} are related by the coordinate transformation\footnote{Another form of the warped AdS$_3$ black hole metric that also reduces to BTZ in the unwarped limit was introduced in Eq.~(2.23) of \cite{Martin:2022duk}.
However, that change of coordinates seems to break down in the extremal case.
On the other hand, it is claimed in Eqs.~(4.26) and (4.27) therein that the BTZ QNMs are recovered in the unwarped limit, contrary to the conclusion we reach at the end of this appendix.
This apparent discrepancy deserves further clarification.}
\begin{align}
    \hat{t}=\frac{L}{W}t\,,\qquad
    \hat{\theta}=-\theta-\frac{t}{W}\,,\qquad
    \hat{r}^2=\frac{\nu^2+3}{4\nu^2}\left(W\nu r-\frac{3}{4}r_+r_-\left(\nu^2-1\right)\right)\,,
\end{align}
with
\begin{subequations}
\begin{align}
    \hat{r}_{\pm}^2&=\frac{\nu^2+3}{4\nu^2}\left(W\nu r_{\pm}-\frac{3}{4}r_+r_-\left(\nu^2-1\right)\right)\,,\\
    W&=(r_++r_-)\nu-\sqrt{r_+r_-\left(\nu^2+3\right)}\,.
\end{align}
\end{subequations}
The constants $H$ and $L$ in the quadratic ensemble control the parameter $\nu$ and the AdS radius $l$ in the metric \eqref{eq:WBTZ}, where we set $l=1$.
These parameters are related through
\begin{eqnarray}
    H^2=-\frac{3\left(\nu^2-1\right)}{2\left(\nu^2+3\right)}\,,\qquad L=\frac{2l}{\sqrt{\nu^2+3}}\,.
\end{eqnarray}

We now compute the parameters of the photon rings (namely, their radii $\tilde{\hat{r}}_\pm$, angular velocities $\hat{\Omega}_\pm$ and Lyapunov exponents $\hat{\gamma}_\pm$) and express them in terms of the parameters of the photon rings of the metric \eqref{eq:WBTZ}: 
\begin{subequations}
\begin{align}
    \tilde{\hat{r}}_\pm^2&=\hat{r}^2(\tilde{r}_\pm)
    =\frac{\hat{r}_+^2+\hat{r}_-^2}{2}\pm\sqrt{1-\frac{1}{2H^2}}\frac{\hat{r}_+^2-\hat{r}_-^2}{2}\,,\\ 
    \hat{\Omega}_\pm&=-\frac{W\tilde{\Omega}_\pm+1}{L}
    =\frac{\sqrt{1-\frac{1}{2H^2}}(\hat{r}_++\hat{r}_-)\mp(\hat{r}_+-\hat{r}_-)}{\sqrt{1-\frac{1}{2H^2}}(\hat{r}_++\hat{r}_-)\pm\left(\hat{r}_+-\hat{r}_-\right)}\,,\\ 
    \hat{\gamma}_{L\pm}&=\frac{W}{L}\gamma_{L\pm}
    =\frac{2}{L^2}\frac{\sqrt{1-\frac{1}{2H^2}}(\hat{r}_+^2-\hat{r}_-^2)}{\sqrt{1-\frac{1}{2H^2}}(\hat{r}_++\hat{r}_-)\pm\left(\hat{r}_+-\hat{r}_-\right)}\,.
\end{align}
\end{subequations}
The BTZ limit $H\to0$ (or $\nu\to1$) of these parameters is
\begin{align} 
    \tilde{\hat{r}}_\pm^2\to\pm\infty\,,\qquad
    \hat{\Omega}_\pm\to1\,,\qquad
    \hat{\gamma}_{L\pm}\to2(\hat{r}_+-\hat{r}_-)\,.
\end{align}
The fate of the inner photon ring in this limit is subtle, since its radius is pushed to $-\infty$ but the domain of $r$ is $\mathbb{R}^+$.
Thus, the inner photon ring must vanish in this limit, while the outer one is pushed to the AdS boundary at spatial infinity.
 
Starting with the eikonal spectrum \eqref{eq:OuterQNMs} of QNMs associated with the outer photon ring,
\begin{align}
    \omega_1=\tilde{\Omega}_+k-i\left(n+\frac{1}{2}\right)\gamma_{L+}\,,\qquad
    \omega_2=i\nu(2n+1)+i\sqrt{3\left(\nu^2-1\right)n(n+1)+\nu^2}\,,
\end{align}
where $n$ is an integer, and by defining new modes via
\begin{align}
    e^{-i\omega t+ik\theta}R(z)=e^{-i\hat{\omega}\hat{t}+i\hat{k}\hat{\theta}}R\left(\frac{\hat{r}^2-\hat{r}_+^2}{\hat{r}^2-\hat{r}_-^2}\right)\,,
\end{align}
we end up in the quadratic ensemble with the QNM frequencies 
\begin{subequations}
\label{eq:NotImaginary}
\begin{align}
    \hat{\omega}_1&=\hat{\Omega}_+\hat{k}-i\left(n+\frac{1}{2}\right)\hat{\gamma}_{L+}\,,\\
    \hat{\omega}_2&=-\hat{k}-i\frac{\left(1-2H^2\right)\left(\hat{r}_++\hat{r}_-\right)}{\sqrt{1+\frac{2}{3}H^2}L}\left[\sqrt{1-2H^2}(2n+1)-\sqrt{1-2H^2[1+4n(n+1)]}  \right]\,.
\end{align}
\end{subequations}

We now take the BTZ limit $H^2\to 0$ to see whether we recover the BTZ spectrum \eqref{eq:SpectrumBTZ}.
To do so correctly, we examine the exact spectrum, rather than its eikonal limit.
For $H=0$, it is
\begin{align}
    \hat{\omega}_1=\hat{k}-i\left(\hat{r}_+-\hat{r}_-\right)\left[2n+1-\sign(\hat{k})\right]\,,\qquad
    \hat{\omega}_2=-\hat{k}-2i\left(\hat{r}_++\hat{r}_-\right)n\,.
\end{align}
We can rewrite the modes $\hat{\omega}_1$ in terms of the outer photon ring parameters as
\begin{align}
    \hat{\omega}_1=\hat{\Omega}_+\hat{k}-i\left(n+\frac{1-\sign(\hat{k})}{2}\right)\hat{\gamma}_{L+}\,.
\end{align}
We see that in the BTZ limit $H\to0$, we do not exactly recover \eqref{eq:SpectrumBTZ}.
This reflects the fact that setting $H=0$, and then computing the QNM spectrum (the computation of \cite{Birmingham:2001hc}) is not equivalent to taking the unwarped limit $H\to0$ of the QNM spectrum we computed in the warped case $H\neq0$.
In other words, the two limits do not commute.

Physically, this phenomenon can be traced back to the fact that the nature of infinity is drastically different depending on whether $H=0$ or not.
In particular, the QNM boundary conditions are not continuously deformed one into another as $H\to0$.
This could be compared, for graviton perturbations instead of the scalar ones considered here, to the determination of boundary conditions in asymptotically WAdS$_3$ spacetimes \cite{Compere:2009zj, Ciambelli:2020shy}.
Despite the fact that for $H=0$ the metrics become asymptotically AdS$_3$, the asymptotic symmetries of WAdS$_3$ do not reduce to the conformal Brown-Henneaux boundary conditions \cite{Brown:1986nw}, but rather to the warped conformal CSS ones \cite{Compere:2013bya}.

\bibliographystyle{JHEP}
\bibliography{references}

\end{document}